%% file: aparna.tex
\documentclass{article}

\input preamble.tex

\usepackage{epsfig}
\usepackage{multicol}
\usepackage{pslatex}
\usepackage{apalike}

\advance\oddsidemargin by -0.45in
\advance\textwidth by 0.9in
\advance\topmargin by -0.6in
\advance\textheight by 1.2in

\begin{document}

\title{Malware Classification with Word Embedding Features}

\author{Aparna Sunil Kale 
\and
Fabio Di Troia
\and 
Mark Stamp\footnote{mark.stamp@sjsu.edu}
}

\maketitle


\abstract{Malware classification is an important and challenging problem in information security. 
Modern malware classification techniques rely on machine learning models that can be 
trained on features such as opcode sequences, API calls, 
and byte $n$-grams, among many others. In this research, we consider opcode
features. We implement hybrid machine 
learning techniques, where we engineer feature vectors by training
hidden Markov models---a technique that we refer to as HMM2Vec---and 
Word2Vec embeddings on these opcode sequences. The resulting HMM2Vec 
and Word2Vec embedding vectors are then used as features for classification algorithms. 
Specifically, we consider support vector machine (SVM), $k$-nearest neighbor ($k$-NN), 
random forest (RF), and convolutional neural network (CNN) classifiers. We conduct substantial 
experiments over a variety of malware families. Our experiments extend well beyond
any previous work in this field.}


\section{Introduction}

Malware is a software that is created with the intent to cause harm to computer data 
or otherwise adversely affect computer systems~\cite{R_1}. Detecting malware can be 
a challenging task, as there exist a wide variety of advanced malware that employ 
various anti-detection techniques.

Some advanced types of malware include polymorphic and metamorphic code.
Such code enables attackers to easily generate vast numbers of new 
malware variants that cannot be detected using standard techniques~\cite{R_1}. 
According to McAfee, some~60 million new malware samples were created 
in the first quarter of the year~2019~\cite{R_2}. Malware detection---and
the closely related problem of malware classification---are inherently challenging 
in such an environment.

The methods used to protect users from malware include signature-based and 
behavioral-based detection techniques. Signature-based detection is 
based on specific patterns found in malware samples. In contrast, behavioral-based detection 
is based on actions performed by code. Signature-based techniques cannot detect
new malware variants, while behavioral-based techniques often results in a 
high false-positive rate. Both of these traditional malware detection strategies
fail when confronted with advanced forms of malware~\cite{R_3}. Therefore, it is essential to 
consider alternative malware detection techniques. 

Modern malware research
often focuses on machine learning, which has shown better performance as compared 
to traditional methods, particularly in the most challenging cases. Machine learning models 
for malware classification can be trained on a wide variety of features, including API calls, 
opcodes sequences, system calls, and control flow graphs, among many others~\cite{R_4}.

In this research, we extract features from several malware families. 
We then train a hidden Markov model (HMM) for each malware sample. 
The converged probability matrices of the HMMs
(the so-called~$B$ matrices) that belong to the same malware family would be
expected to have similar characteristics. These matrices from our converged HMMs 
are converted to feature vectors, in a process that we refer to as HMM2Vec.
The resulting HMM2Vec vectors are used as features 
in the following classification algorithms: $k$-nearest neighbors ($k$NN), 
support vector machines (SVM), random forest (RF), and convolutional neural networks (CNN);
we refer to these hybrid techniques as 
HMM2Vec-$k$NN, HMM2Vec-SVM, HMM2Vec-RF, and HMM2Vec-CNN, 
respectively. Similarly, we consider Word2Vec encodings in place of the HMM2Vec
embeddings, which gives rise to the corresponding hybrid models of
Word2Vec-$k$NN, Word2Vec-SVM, Word2Vec-RF, and Word2Vec-CNN.

For each of these hybrid techniques, extensive malware classification experiments
are conducted over a set of seven challenging malware families. Again,
our experiments are based on engineered features derived from opcode sequences.

The remainder of this paper is organized as follows.
In Section~\ref{chap:background} we provide a discussion of relevant background topics, 
with a focus on the machine learning techniques employed
in this research. We also provide a selective survey of related work. 
Section~\ref{chap:implementation} covers the novel 
hybrid machine learning techniques that are the focus of this research. In this 
section, we also provide information on the dataset that we have used. 
Section~\ref{chap:results} gives our experimental 
results and analysis. Finally, 
Section~\ref{chap:conclusion} summarizes our results and includes a
discussion of possible directions for future work.

\section{Background}\label{chap:background}

In this section, we introduce the machine learning models used in this research.
We also provide a selective survey of relevant previous work.

\subsection{Machine Learning Techniques}

A wide variety of machine learning techniques are considered in this research.
We train hidden Markov models and generate Word2Vec embeddings, which are
subsequently used as features in various classification algorithms. The classification
algorithms considered are random forest, $k$-nearest neighbor, support vector machines,
and convolutional neural networks. Due to space limitations, each of these 
techniques is introduced only very briefly in this section.

\subsubsection{Hidden Markov Models}

A hidden Markov model (HMM) is a probabilistic machine learning algorithm
that can be used for pattern matching applications in such diverse areas
as speech recognition~\cite{Rabiner},
human activity detection~\cite{human_activity}, and protein sequencing~\cite{protein}.
HMMs have also proven useful for malware analysis.
 

A discrete HMM is defined as~$\lambda=(A,B,\pi)$, where~$A$ is the state transition
matrix for the underlying Markov process, $B$ contains probability distributions that
relate the hidden states to the observations, and~$\pi$ is the initial state distribution.
All three of these matrices are row stochastic.

In this research, we focus on the~$B$ matrix of trained HMMs. These matrices
can be viewed as representing crucial statistical properties of the observation sequences
that were used to train the HMM. Using these~$B$ matrices as input to classifiers
is an advanced form of feature engineering, whereby information in the 
original features is distilled into a potentially more informative form by the 
trained HMM. We refer to the process of deriving these HMM-based feature
vectors as HMM2Vec. We have more to say about 
generating HMM2Vec features from HMMs in Section~\ref{sect:hmm_based}.
 
\subsubsection{Word2Vec}

Word2Vec has recently gained considerable popularity in 
natural language processing (NLP)~\cite{R_5}.
This word embedding technique is based on a shallow neural network, with the weights
of the trained model serving as embedding vectors---the trained model itself serves no
other purpose. These embedding vectors capture significant relationships 
between words in the training set. Word2Vec can also
be used beyond the NLP context to model relationships between 
more general features or observations.

When training a Word2Vec model, we must specify the desired
vector length, which we denote as~$N$. Another key parameter
is the window length~$W$, which represents the width of a sliding
window that is used to extract training samples from the data.
Algebraic properties hold for Word2Vec
embeddings; see~\cite{w2v} for further information. 

In this research, we train Word2Vec models on opcode sequences.
The resulting embedding vectors are used as feature
vectors for several different classifiers. Analogous to the HMM feature vectors discussed
in the previous section, these Word2Vec embeddings serve as
engineered features that may be more informative than the raw opcode
sequences.

\subsubsection{Random Forest}

Random forest (RF) is a class of supervised machine learning techniques. 
A random forest is based on decision trees, which are one of the
simplest and most intuitive ``learning'' techniques available. The primary drawback to
a simple decision trees is that it tends to overfit---in effect, the decision
tree ``memorizes'' the training data, rather than learning from the it.
A random forest overcomes this limitation by a process known as ``bagging,''
whereby a collection of decision trees are trained, each using a subset of the
available data and features~\cite{R_18}. 


\subsubsection{$k$-Nearest Neighbors}

Perhaps the simplest learning algorithm possible is $k$-nearest neighbors~($k$NN). 
In this technique, 
there is no explicit training phase, and in the testing phase, a sample is
classified simply based on the nearest neighbors in the training set.
This is a lazy learning technique, in the sense that all computation
is deferred to the classification phase. The parameter~$k$ specifies the number
of neighbors used for classification. Small values of~$k$ tend to results
in highly irregular decision boundaries, which is a hallmark of overfitting.
 
\subsubsection{Support Vector Machine}

Support vector machines (SVM) are popular supervised learning algorithms~\cite{R_17}
that have found widespread use in malware analysis~\cite{R_10}. 
A key concept behind SVMs is a separating hyperplane that
maximizes the margin, which is the minimum distance between the classes.
In addition, the so-called kernel trick introduces nonlinearity into the process, with minimal
computational overhead.



Several popular nonlinear kernels are used in SVMs.
In this research, we experiment with linear kernels and nonlinear
radial basis function (RBF) kernels.

\subsubsection{Convolutional Neural Network}

Neural networks are a large and diverse class of learning algorithms
that are loosely modeled on structures of the brain.
A deep neural network (DNN) is a neural network with multiple hidden
layers---such networks are state of the art for many learning problems.
Convolutional neural networks (CNN) are DNNs that are 
optimized for image analysis, but have proven effective in many other
problem domains. 
The architecture of a CNN consists of hidden layers, along with input and output layers. 
The hidden layers of a CNN typically include convolutional layers, 
pooling layers, and a fully connected output layer~\cite{R_5}.
 
\subsection{Selective Survey of Related Work}

Machine learning has been widely used in malware research. 
This section introduces representative examples from the literature
that are related to the work considered in this paper.
 
In~\cite{R_11}, the authors consider features based on API calls. 
Other research considers features such as opcodes,
system calls, control flow graphs, and byte sequences for malware detection~\cite{R_10,R_9}. 

As discussed in~\cite{R_11}, features obtained via static analysis (e.g., opcode sequences) 
results in more efficient and faster techniques as compared to those that rely on features that
require dynamic analysis (e.g., API calls). However, dynamic analysis often provides
a more accurate reflection of malware, and obfuscation techniques are generally
less effective for dynamic features.
 
 
The literature is replete with hybrid machine learning techniques for malware classification. 
For example, in~\cite{R_13}, the author proposes a hybrid machine learning technique 
that uses HMM matrices as the input to a convolutional neural network to classify malware families. 
Researchers in~\cite{R_13} use SVMs to classify trained HMMs. In~\cite{Xception}, 
the authors consider an ensemble model that combines predictions from 
\texttt{asm} and \texttt{exe}, files together---the predictions are stacked and fed 
to a neural network for classification. 
In~\cite{R_12}, the authors use 
Word2Vec to generate embeddings from machine instructions. Moreover, they 
propose a proof of concept model to train a convolutional neural network based 
on the Word2Vec embeddings.

The research in this paper builds on
the work in~\cite{R_12,R_13,R_11}. 
We propose hybrid machine learning techniques 
for malware classification using HMM2Vec 
and Word2Vec engineered features which are derived from 
opcode sequences. Four different classifiers are considered, 
giving us a total of eight distinct experiments that we refer to as
HMM2Vec-$k$NN, HMM2Vec-SVM, HMM2Vec-RF, HMM2Vec-CNN,
Word2Vec-$k$NN, Word2Vec-SVM, Word2Vec-RF, and Word2Vec-CNN.
As far as the authors are aware, only one of these eight combinations,
namely, Word2Vec-CNN, has been considered in
previous work. Moreover, we experiment with a much wider array
of window sizes and vector lengths for our Word2Vec models
as compared to prior related work.
In the next section, we discuss our eight proposed 
techniques in detail.

\section{Implementation}\label{chap:implementation}

In this section, we first give information about the dataset used in this research.
Then we discuss the various hybrid machine learning techniques that are the focus
of the experiments reported in Section~\ref{chap:results}.

\subsection{Dataset}\label{sect:data}

\begin{figure}[!htb]
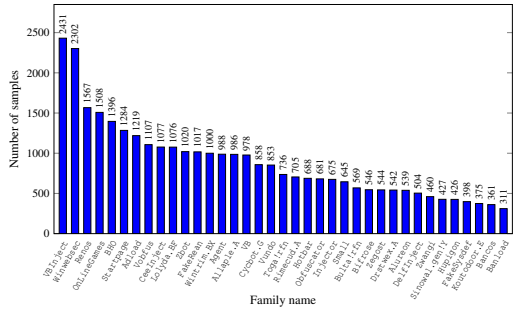

\centering
\input figures/samples.tex
\caption{Families in dataset with at least~300 samples}\label{fig:/bigger_dataset_histogram}
\end{figure}

The raw dataset used for our experiments includes~2793 malware families
with one or more samples per family~\cite{sKim}. 
Figure~\ref{fig:/bigger_dataset_histogram} 
lists all of the families in our dataset that have more than~300 samples.

We selected seven of the families listed in Figure~\ref{fig:/bigger_dataset_histogram} 
that have more than~1000 samples, and randomly selected~1000
samples of each type, giving us a total of~7000 samples.
The following seven families were selected for this research.

\begin{description}
\item[BHO] can perform a wide variety of malicious actions, 
as specified by an attacker~\cite{R_19}. 
\item[CeeInject] is designed to conceal itself from detection, 
and hence various families use it as a shield to prevent detection. 
For example, CeeInject can obfuscate a bitcoin mining client,
which might have been installed on a system without the user's 
knowledge or consent~\cite{R_22}.	
\item[FakeRean] pretends to scan the system, notifies the user of nonexistent issues, 
and asks the user to pay to clean the system~\cite{R_23}. 
\item[OnLineGames] steals login information and captures user 
keystroke activity~\cite{R_21}. 
\item[Renos] will claim that the system has spyware and ask for a payment to 
remove the nonexistent spyware~\cite{R_20}. 
\item[Vobfus] is a family that downloads other malware onto a user's computer
and makes changes to the device configuration that cannot be restored 
by simply removing the downloaded malware~\cite{R_24}.
\item[Winwebsec] is a trojan that presents itself as antivirus software---it displays
misleading messages stating that the device has been 
infected and attempts to persuade the user to pay a fee to free the 
system of malware~\cite{R_15}.
\end{description}

For each sample, we train an HMM and a Word2Vec model using opcode sequences. 
The raw dataset consists of \texttt{exe} files, and hence we first extract the
mnemonic opcode sequence from each malware sample. 
We use \texttt{objdump} to generate \texttt{asm} files from which we
extract opcode sequences. For each opcode sequence
we retain the~$M$ most frequent opcodes and remove all others. We
experiment with the~$M$ most frequent opcodes for~$M\in\{20,31,40\}$, 
where ``most frequent'' is based on the opcode distribution over the entire dataset. 
The number of hidden states in each HMM was chosen to be~$N=2$, 
and the number of output symbols is given by~$M$. 
For the Word2Vec models, we experiment with additional parameters.

\begin{figure}[!htb]
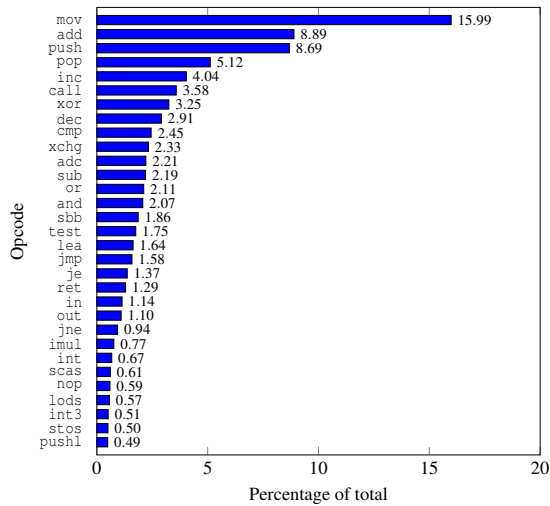

\centering
\input figures/opcodes.tex
\caption{The~31 most frequent opcodes}\label{fig:/raw_dataset_31opcode}
\end{figure}

Experiments involving~$M\in\{20, 31,40\}$ are discussed at the start
of Section~\ref{chap:results}.
Based on the results of such experiments, we selected~$M=31$ for
all subsequent HMM2Vec and Word2Vec experiments. 
The~31 most frequent opcodes are
listed in Figure~\ref{fig:/raw_dataset_31opcode},
along with the percentage of the total that each opcode
represents.

From the numbers in Figure~\ref{fig:/raw_dataset_31opcode},
we see that any opcode outside of the top~31 accounts for less
that 0.5\%\ of the total opcodes. Since we are considering
statistical-based feature engineering techniques, these omitted
opcodes are highly unlikely to affect the results to any significant degree.

\subsection{Hybrid Classification Techniques}

In this section, we discuss the hybrid machine learning models 
that are the basis for the research in this paper.
Specifically, we consider
HMM2Vec-SVM, HMM2Vec-RF, HMM2Vec-$k$NN, and HMM2Vec-CNN.
We then briefly discuss the analogous Word2Vec techniques,  namely,
Word2Vec-SVM, Word2Vec-RF, Word2Vec-$k$NN, and Word2Vec-CNN.

To train our hidden Markov models, we use
the \texttt{hmmlearn} library~\cite{R_hmmlearn},
and we select the best HMM based on multiple random restarts.
For all remaining machine learning techniques, except for CNNs,
we used \texttt{sklearn}~\cite{scikit-learn}.
To train our CNN models, we use the Keras library~\cite{R_25}.

\subsubsection{HMM Hybrid Techniques}

For our HMM2Vec-SVM hybrid technique, we first train an HMM for each sample,
using the extracted opcode sequence as the training data.
Then we use an SVM to classify the samples, based on the~$B$
matrices of the converged HMMs.
Each converged~$B$ matrix is vectorized by simple concatenating the rows.
Since~$N=2$ is the number of hidden states and~$M$ is the number of 
distinct opcodes in the observation sequence, 
each~$B$ matrix is~$N\times M$. Consequently, the
resulting engineered feature vectors are all of length~$NM$. 
When training the SVM,
we experiment with various hyperparameters and kernel functions.

Our HMM2Vec-RF, HMM2Vec-$k$NN, and HMM2Vec-CNN techniques are 
analogous to the HMM2Vec-SVM hybrid technique.
For the HMM2Vec-CNN, we use a one-dimensional CNN.
In each case, we tune the relevant parameters. 

\subsubsection{Word2Vec Hybrid Techniques}

As mentioned above, Word2Vec is typically trained on a series of words, 
which are derived from sentences in a natural language. In our research, 
the sequence of opcodes from a malware executable is treated as a stream of ``words.''
Analogous to our HMM2Vec experiments, we concatenate the Word2Vec
embeddings to obtain a vector of length~$NM$, where~$M$ is the number
of distinct opcodes in the training set and~$N$ is the length of the 
embedding vectors.

Once we have trained the Word2Vec models to obtain the 
engineered feature vectors, the classification process for each of
Word2Vec-SVM, Word2Vec-RF, Word2Vec-CNN, and Word2Vec-$k$NN
is analogous to that for the corresponding HMM-based technique.
As with the HMM classification techniques, we tune the parameters in each case.

\section{Experiments and Results}\label{chap:results}

In this section, we present the results of several hybrid machine learning experiments
for malware classification.
As discussed above, these experiments are based on opcode sequences,
with feature engineering involving HMM and Word2Vec models.
We consider four classifiers, giving us a total of eight different
experiments, which we denote as
HMM2Vec-SVM, HMM2Vec-RF, HMM2Vec-$k$NN, HMM2Vec-CNN, 
Word2Vec-SVM, Word2Vec-RF, Word2Vec-$k$NN, and Word2Vec-CNN. 

Before discussing our hybrid multiclass results, we first consider
binary classification experiments using different numbers of opcodes.
The purpose of these experiments is to determine the number of opcodes
to use in our subsequent multiclass experiments.

\subsection{Binary Classification} 

In this section, we classify samples from the
Winwebsec and Fakerean malware families, both of which are examples of rogue 
security software that claim to be antivirus tools. We compare the accuracies 
when using the~$M$ most frequent opcodes, for~$M\in\{ 20, 31, 40\}$. 

\begin{figure}[!htb]
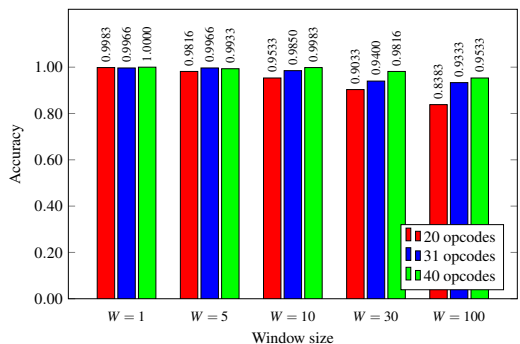

\centering
\input figures/w2v_svm_2.tex
\caption{Binary classification using Word2Vec-SVM model (Winwebsec vs Fakerean)} 
\label{fig:/opcode_variety}
\end{figure}

For each of these binary classification experiments, we generate a Word2Vec 
model for each sample in both families, using a vector size of~$N=2$ and window 
sizes of~$W\in\{1, 5, 10, 30, 100\}$. Thus, we conduct~15 distinct experiments,
each involving~2000 labeled samples. In each case, we use a~70-30
training-testing split. The results of these experiments are summarized in
Figure~\ref{fig:/opcode_variety}. 

From Figure~\ref{fig:/opcode_variety}, we see that
good results are obtained 
for window size~$W=5$ and~31 or~40 opcodes.
Both of these cases yield an accuracy in excess of~99\%.
But, the improvement when using~40 opcodes over~31
opcodes is relatively small, and with~40 opcodes,
feature extraction and training times are greater.
Therefore, in all of the multiclass experiments
discussed in the next sections, we use~31 opcodes.

\subsection{HMM2Vec Multiclass Experiments}\label{sect:hmm_based}

For all of our multiclass experiments, we
consider the seven malware families that are discussed in Section~\ref{sect:data},
namely,
BHO~\cite{R_19}, 
Ceeinject~\cite{R_22}, 
Fakerean~\cite{R_23}, 
OnLineGames~\cite{R_21}, 
Renos~\cite{R_20}, 
Vobfus~\cite{R_24},
and 
Winwebsec~\cite{R_15}.
We extracted opcodes from~50 malware families
and use the~31 most frequent to train HMMs for each sample in each of the seven 
families under consideration.
For all HMMs, the number of hidden states is selected to be~$N=2$.
Since we are considering~31 distinct opcodes, we have~$M=31$,
giving us engineered HMM2Vec feature vectors of length~62. 

As mentioned above,
we train HMMs using the \texttt{hmmlearn} library~\cite{R_hmmlearn}
and we select the highest scoring model 
based on multiple random restarts.
The precise number of random restarts is determined by the
length of the opcode sequence---for shorter sequences in the range 
of~1000 to~5000 opcodes, we use~100 restarts; otherwise
we select the best model based on~50 random restarts. 
The~$B$ matrix of the highest-scoring model is then converted 
to a one-dimensional vector. 

To obtain the HMM2Vec features, 
we convert the~$B$ matrix of a trained HMM into vector form.
A subtle point that arises in this conversion process is that
the order of the hidden states in the~$B$ matrix need not be consistent
across different models. Since we only have~$N=2$ hidden states
in our experiments, this means that
the order of the rows of the corresponding~$B$ 
matrices may not agree between different models. To account for this
possibility, we
determine the hidden state that has the highest probability with respect to the
\texttt{mov} opcode and we deem this to be the first half of the
HMM2Vec feature vector, with the other row of the~$B$ matrix
being the second half of the vector. Since \texttt{mov} is by far the most frequent
opcode, this will yield a consistent ordering of the hidden states.

\subsubsection{HMM2Vec-SVM}

Table~\ref{tab:hmm_svm_gridsearch} gives the results of a grid search over
various parameters and kernel functions. As with all of our multiclass
experiments, we use a~70-30 split of the data for training and testing.
For the multiclass SVM, we use a one-versus-other technique.
From the results in Table~\ref{tab:hmm_svm_gridsearch}, we see that the
RBF kernel performs poorly, while the linear kernel yields consistently strong 
results. Our best results are obtained using a linear kernel with~$C=100$
and~$C=1000$.

\begin{table}[!htb]
\caption{HMM2Vec-SVM accuracies}\label{tab:hmm_svm_gridsearch}
\vglue 0.1in
\centering
\begin{tabular}{ccc|c}\midrule\midrule
\multirow{2}{*}{\textbf{Kernel}} 
			& \multicolumn{2}{c|}{\textbf{Parameters}}
			& \multirow{2}{*}{\textbf{Accuracy}} \\ \cmidrule{2-3}
                         & $C$ & $\gamma$ \\ \midrule
linear & \z\z\z1 & N/A & 0.83 \\ 
linear & \z\z10 & N/A & 0.87 \\  
linear & \z100 & N/A & 0.88 \\ 
linear & 1000 & N/A & 0.88 \\ \midrule 
RBF & \z\z\z1 & 0.001\z & 0.13 \\ 
RBF & \z\z\z1 & 0.0001  & 0.13 \\ 
RBF & \z\z10 & 0.001\z & 0.42 \\ 
RBF & \z\z10 & 0.0001 & 0.13 \\ 
RBF & \z100 & 0.001\z & 0.69 \\ 
RBF & \z100 & 0.0001 & 0.34 \\ 
RBF & 1000 & 0.001\z & 0.83 \\ 
RBF & 1000 & 0.0001 & 0.70 \\ \midrule\midrule
\end{tabular}
\end{table}

%

\begin{figure}[!htb]
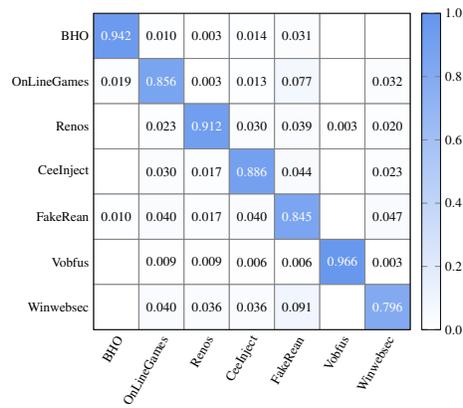

\centering
\input figures/conf_hmm_svm.tex
\caption{Confusion matrix for HMM2Vec-SVM with linear kernel} 
\label{fig:/hmm_svm_linear}
\end{figure}

Figure~\ref{fig:/hmm_svm_linear} gives the 
confusion matrix for our HMM2Vec-SVM experiment, based on
a linear kernel with~$C=100$.
We see that BHO and Vobfus are classified with the highest 
accuracies of~94.2\%\ and~96.6\%, respectively. On the other
hand, Winwebsec and Fakerean are the most challenging, with~9\%\ 
and~7\%\ misclassification rates, respectively. We also note that 
OnLineGames samples are frequently misclassified as Fakerean.

\subsubsection{HMM2Vec-$k$NN}

\begin{figure}[!htb]
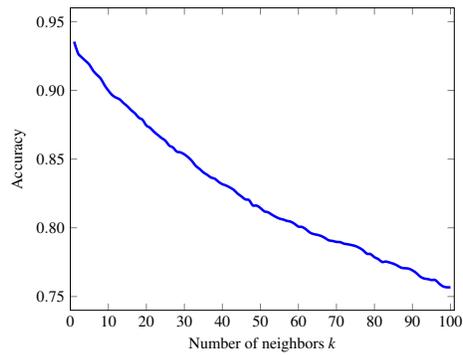

\centering
\input figures/hmm_knn2.tex
\caption{HMM2Vec-$k$NN accuracy as a function of~$k$} 
\label{fig:/knn_cv_B}
\end{figure}

Recall that in~$k$NN, the parameter~$k$ is the number of 
neighbors that are used to classify samples. 
We experimented with $k$NN classifiers using our engineered HMM2Vec 
features for each~$k\in\{1,2,3,\ldots,50\}$. Figure~\ref{fig:/knn_cv_B} 
gives the resulting multiclass accuracy for these HMM2Vec-$k$NN 
experiments as a function of~$k$.

From Figure~\ref{fig:/knn_cv_B}, we see that as the accuracy declines
as~$k$ increases. However, 
small values of~$k$ result in a highly irregular decision boundary, 
which is a sign of overfitting. As a general rule, we should choose~$k\approx\sqrt{S}$,
where~$S$ is the number of training samples. For our experiment, this
gives us~$k=70$, for which we obtain an accuracy of about~79\%.

\subsubsection{HMM2Vec-RF}

There are many hyperparameters to consider when training a random forest.
Using our HMM engineered features, we performed a randomized search and 
obtained the best results with the parameter in Table~\ref{tab:hmm_rf_gridsearchcv}.

\begin{table}[!htb]
\caption{Randomized search parameters for HMM2Vec-RF}
\label{tab:hmm_rf_gridsearchcv}
\vglue 0.1in
\centering
\begin{tabular}{cc}\midrule\midrule
 \textbf{Hyperparameter} & \textbf{Value} \\ \midrule
 $n$-estimators & 1000\\ 
 min samples split & \z\z\z2\\  
 min samples leaf & \z\z\z1\\  
 max features & auto\\
 max depth & \z\z50\\
 bootstrap &false\\ \midrule\midrule 
\end{tabular}
\end{table}

\begin{figure}[!htb]
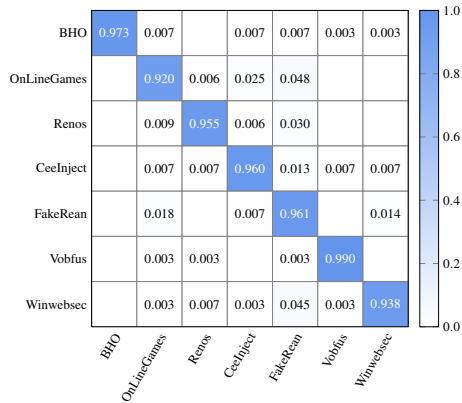

\centering
\input figures/conf_hmm_rf_grid.tex
\caption{Confusion matrix for HMM2Vec-RF using grid parameters}  
\label{fig:/hmm_rf_grid}
\end{figure}

Using the hyperparameters in Table~\ref{tab:hmm_rf_gridsearchcv}, 
our HMM2Vec-RF classifier achieves an overall accuracy of~96\%.
In Figure~\ref{fig:/hmm_rf_grid}, we give the results of this experiment
in the form of a confusion matrix.
From this confusion matrix, we see that
BHO and Vobfus are classified with high accuracies 
of~97\%\ and 99\%, respectively. The misclassifications between
OnLineGames and Fakerean are reduced, as compared to the SVM
classifier considered above, as are the misclassifications between Winwebsec 
and Fakerean.

\subsubsection{HMM2Vec-CNN}

Next, we consider classification based on CNNs.
There are numerous possible configurations and many hyperparameters in such models.
Due to the fact that our feature vectors are one-dimensional,
we use one-dimensional CNNs. 

We split the data into~80\%\ training, 10\%\ validation, 
and~10\%\ testing. With this split, we have~5600 training samples, 
700 validation samples, and~700 testing samples. We train each 
model using the rectified linear unit (ReLU) activation function 
and~200 epochs. To construct these models, we used the Keras library~\cite{R_25}. 

For our first set of experiments, we train CNNs using one input layer of 
dimension~200, a hidden layer with~500 neurons, and
our output layer has seven neurons, since we have seven classes. 
We use a mean squared error (MSE) loss function.

Using stochastic gradient descent (SDG) as the optimizer,
we obtained an accuracy of about~50\%.
Switching to the Adam~ optimizer~\cite{R_26}, 
we achieve a training accuracy of~97\%\ and a testing accuracy of~92\%. 
Consequently, we use Adam for all further experiments.

We train a CNN with two hidden layers, one input layer, 
and an output layer, with~20, 200, and seven neurons, respectively. 
In this case, using categorical cross-entropy (CC) as the loss function,
we achieved a testing accuracy of~88\%, but the model showed a~40\%\ loss. 

Next, we expand the hidden layer to~500 neurons
and perform a grid search to identify the best hyperparameters. 
We experimented with various loss functions, 
we use~200 neurons in the input layer, one hidden layer with~500 neurons,
ReLU activation functions, and an output layer with seven neurons,
followed by a softmax activation. In this setup, we achieved a testing accuracy 
of~93.8\%\ using the CC loss function. However, the training accuracy 
reached~100\%, which indicates overfitting. 
%

There are several possible ways to mitigate overfitting---we employ regularization based on 
a dropout layer. 
Intuitively, when a neuron is dropped at a particular iteration, it forces other neurons
to become active, which reduces the tendency of some neurons 
to dominate during training. 
By spreading the training over more neurons, we
reduce the tendency of the model to overfit the data.


When we set the dropout rate to~0.5, we achieve a testing accuracy 
of~94.2\%\ with a training accuracy of~98\%. In this case, 
we have eliminating the overfitting that was observed in
our previous models.
%

\subsection{Word2Vec Multiclass Experiments}

The experiments in this section are analogous to the HMM2Vec
experiments in Section~\ref{sect:hmm_based}.
However, Word2Vec includes more parameters
that we can easily adjust, as compared to HMM2Vec,
and hence we experiment with these parameters.
Specifically, for our Word2Vec models, we experiment with
different window sizes~$W$
and different lengths~$N$ of the embedding vectors. 
Since we are considering feature vectors with~31 
distinct opcodes, for the~$N=2$ case, we will have
Word2Vec engineered feature vectors of length~62,
which is the same size as the HMM2Vec feature vectors
considered above. However, for~$N>2$, we have larger
feature vectors. Also, the window size allows us to
consider additional context in Word2Vec models,
as compared to our HMM2Vec features.

\subsubsection{Word2Vec-SVM}

Here, we generate feature vectors using Word2Vec,
and apply an SVM classifier. As mentioned above, Word2Vec 
gives us the flexibility to choose the vector embedding
and window sizes, and hence we experiment
with these parameters.
As in all of the multiclass cases, we consider~1000 malware 
samples from each of seven families.
In all cases,
we split the input data~70-30 for training and testing.
For the SVM experiments, we use a one-versus-other technique.

As with our HMM2Vec-SVM experiments, we first perform a grid search 
over the parameters for linear and RBF kernels. For these experiments,
we use vectors size of~$N=2$ and a window of size~$W=30$.
Table~\ref{tab:wv_svm_gridsearch} summarizes the results
of these experiments. We observed that the RBF kernel 
achieves the highest accuracy.

\begin{table}[!htb]
\caption{Word2Vec-SVM grid search accuracies ($N=2$ and $W=30$)}
\label{tab:wv_svm_gridsearch}
\vglue 0.1in
\centering
\begin{tabular}{ccc|c}\midrule\midrule
\multirow{2}{*}{\textbf{Kernel}} 
			& \multicolumn{2}{c|}{\textbf{Parameters}}
			& \multirow{2}{*}{\textbf{Accuracy}} \\ \cmidrule{2-3}
                         & $C$ & $\gamma$ \\ \midrule
linear & \z\z\z1 & N/A & 0.86 \\ 
linear & \z\z10 & N/A & 0.85 \\  
linear & \z100 & N/A & 0.85 \\ 
linear & 1000 & N/A & 0.85\\ \midrule 
RBF & \z\z\z1 & 0.001\z & 0.87 \\ 
RBF & \z\z\z1 & 0.0001  & 0.70 \\ 
RBF & \z\z10 & 0.001\z & 0.91 \\ 
RBF & \z\z10 & 0.0001 & 0.84 \\ 
RBF & \z100 & 0.001\z & 0.92 \\ 
RBF & \z100 & 0.0001 & 0.88 \\ 
RBF & 1000 & 0.001\z & 0.92 \\ 
RBF & 1000 & 0.0001 & 0.90 \\ \midrule\midrule
\end{tabular}
\end{table}

For our next set of Word2Vec-SVM experiments, 
we consider a linear kernel.
For the Word2Vec features,
we use vector lengths~$M\in\{2, 31, 100\}$
and windows of size~$W\in\{1, 5, 10, 30, 100\}$,
giving us a total of~15 distinct Word2Vec-SVM experiments using linear kernels.

\begin{figure*}[!htb]
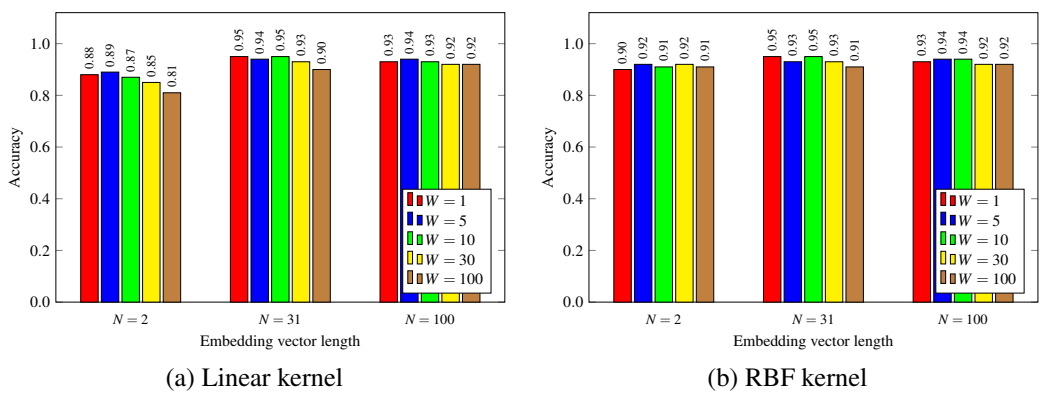

\centering
\begin{tabular}{cc}
\input figures/w2v_svm_linear.tex
& 
\input figures/w2v_svm_rbf.tex
\\
(a) Linear kernel
& 
(b) RBF kernel
\end{tabular}
\caption{Word2Vec-SVM experiments}
\label{fig:/wv_svm_bar_group}
\end{figure*}

The results of these Word2Vec-SVM experiments are summarized in the 
form of a bar graph in Figure~\ref{fig:/wv_svm_bar_group}~(a).
Note that our best accuracy of~95\%\ for the linear
kernel was achieved with input vectors of size~$N=31$ and,
perhaps surprisingly, a window of size~$W=1$. 
These results show that the accuracies significantly improve 
for embedding vector sizes~$N>2$. 


Next, we consider the RBF kernel in more detail.
Based on the results in Table~\ref{tab:wv_svm_gridsearch},
we select~$C=1000$ and~$\gamma=0.001$. 
We generate Word2Vec vectors of sizes~$N\in\{2, 31, 100\}$ and 
window sizes~$W\in\{1, 5, 10, 30, 100\}$. 
The results of these~15 experiments 
are summarized in Figure~\ref{fig:/wv_svm_bar_group}~(b).
In this case, we achieve a best accuracy of~95\%\ with
a vector length of~$N=31$ and a window size of
either~$W=1$ or~$W=10$. 
Note that the results improve when
the vector size~$N$ is increased from~2 to~31,
but the accuracy does not improve for~$N=100$.


\subsubsection{Word2Vec-$k$NN}

\begin{figure}[!htb]
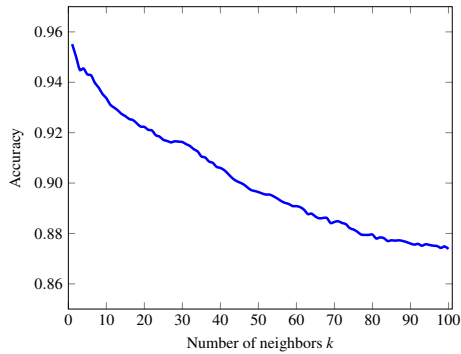

\centering
\input figures/w2v_knn2.tex
\caption{Word2Vec-$k$NN accuracy ($N=100$ and $W=1$)} 
\label{fig:/wv_knn_v100w1}
\end{figure}

For our Word2Vec-$k$NN experiments, we again consider the 15 cases
given by vector lengths~$N\in\{2, 31, 100\}$ and 
window sizes~$W\in\{1, 5, 10, 30, 100\}$.
In each case, we consider~$k\in\{1, 2, 3,\ldots, 100\}$. 
We find that for all cases with vectors with sizes~$N\in\{2, 31, 100\}$ 
and window sizes~$W\in\{1, 5, 10, 30, 100\}$, we achieve about~94\%\  classification 
accuracy. In Figure~\ref{fig:/wv_knn_v100w1} we give a line graph 
for~Word2Vec-$k$NN accuracy as a function of~$k$.
As in our HMM2Vec-$k$NN experiments,
to avoid overfitting, we choose~$k=70$,
which in this case gives us an accuracy of about~89\%.

\subsubsection{Word2Vec-RF}

\begin{figure}[!htb]
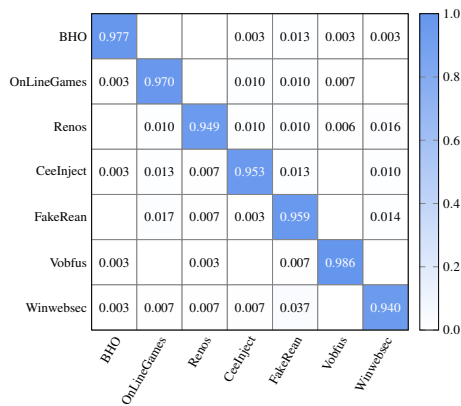

\centering
\input figures/conf_w2v_rf.tex
\caption{Confusion matrix for Word2Vec-RF} 
\label{fig:/wv_rf_v100w30}
\end{figure}

In this set of experiments, we consider the same~15 combinations of
Word2Vec vector sizes and window sizes as in the previous experiments in
this section. In each case, the number of trees in the random
forest is set to~1000. 
We find that the best result for Word2Vec-RF 
occurs with a vector size of~$N=100$ and a window size of~$W=30$,
in which case we achieve an accuracy of~96.2\%. 
The confusion matrix for this case is given in Figure~\ref{fig:/wv_rf_v100w30}. 
The worst misclassification is that Winwebsec is misclassified 
as Fakerean for a mere~3\%\ of the samples tested.

We also conduct experiments on the RF parameters, using a
Word2Vec vector size of~$N=100$ and a window size of~$W=30$. 
Table~\ref{tab:wv_rf_gridsearchcv} lists the best parameters obtained 
based on a grid search. With these parameters, we
obtain an accuracy of~93.17\%.

\begin{table}[!htb]
\caption{Randomized search parameters for Word2Vec-RF}
\label{tab:wv_rf_gridsearchcv}
\vglue 0.1in
\centering
\begin{tabular}{cc}\midrule\midrule
 Hyperparameter & Value \\ \midrule
 $n$-estimators & 1400\\ 
 min samples split & \z\z\z2\\ 
 min samples leaf & \z\z\z1\\  
 max features & auto\\  
 max depth & \z\z40\\ 
 bootstrap &false\\ \midrule\midrule 
\end{tabular}
\end{table}

\subsubsection{Word2Vec-CNN}

Using the same parameters as in the previous Word2Vec experiments, that is,
vector lengths~$N\in\{2, 31, 100\}$ and window sizes~$W\in\{1, 5, 10, 30, 100\}$,
we consider the same CNN architectures as in the HMM2Vec-CNN
experiments, above.

\begin{figure}[!htb]
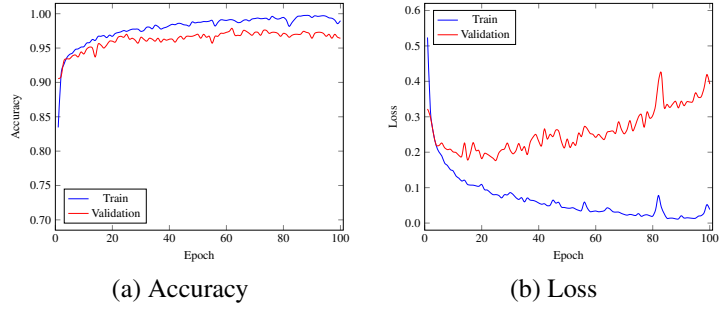

\centering
\begin{tabular}{cc}
\input figures/w2v_dnn_accuracy_overfit.tex
&
\input figures/w2v_dnn_loss_overfit.tex
\\
(a) Accuracy 
&
(b) Loss
\end{tabular}
\caption{Word2Vec-CNN overfittling}\label{fig:/word2vec_dnn_overfit}
\end{figure}

Figure~\ref{fig:/word2vec_dnn_overfit}~(a) and~(b) give model accuracy and loss,
respectively, for an experiments with~$N=31$ and~$W=1$, where we
have trained for~200 epochs. In this case,
we achieve~95\%\ testing accuracy, but validation loss increases dramatically,
which is a clear indication of overfitting. 

\begin{figure}[!htb]
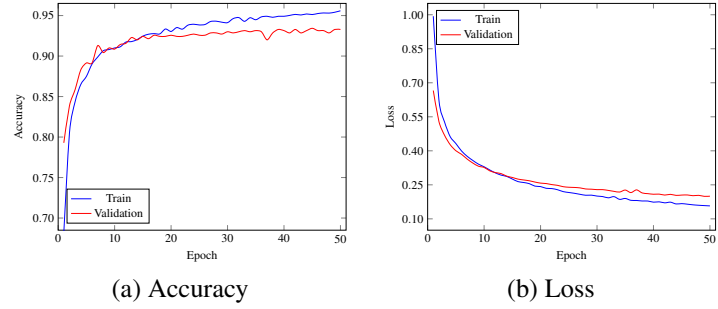

\centering
\begin{tabular}{cc}
\input figures/w2v_dnn_accuracy.tex
&
\input figures/w2v_dnn_loss.tex
\\
(a) Accuracy
&
(b) Loss
\end{tabular}
\caption{Model accuracy and loss for Word2Vec-CNN}\label{fig:/word2vec_dnn_v31w1}
\end{figure}

To deal with the overfitting that is evident in Figure~\ref{fig:/word2vec_dnn_overfit},
we reduce the number of epochs and we tune the learning rate. 
Specifically, we reduce the number of epochs to~50, we 
set the learning rate to~0.0001, and we let~$\beta_1=0.9$ 
and~$\beta_2=0.999$, as per the suggestions in~\cite{R_26}. 
Figure~\ref{fig:/word2vec_dnn_v31w1}~(a) and~(b) give model accuracy and loss, 
respectively, for an experiment based on this selection of parameters. 
In this case, we achieve~94\%\ testing accuracy, 
and the loss is reduced significantly. The loss has been reduced,
and there is no indication of overfitting in this
improved model. 

\begin{figure}[!htb]
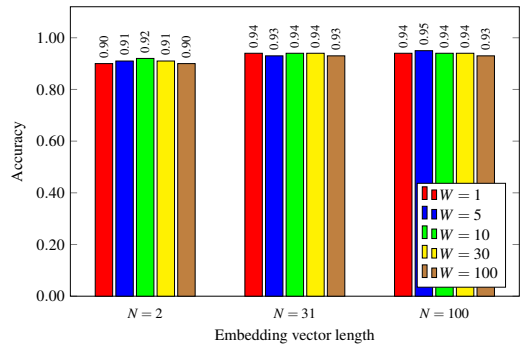

\centering 
\input figures/w2v_dnn.tex
\caption{Accuracies for Word2Vec-CNN experiments} 
\label{fig:/word2vec_dnn_all}
\end{figure}

Figure~\ref{fig:/word2vec_dnn_all} summarized the~15 experiments we 
conducted using Word2Vec-CNN. For these experiment, 
as we increase the window size, generally
we must decrease the number of epochs to keep 
the model loss within acceptable bounds. 

\begin{figure}[!htb]
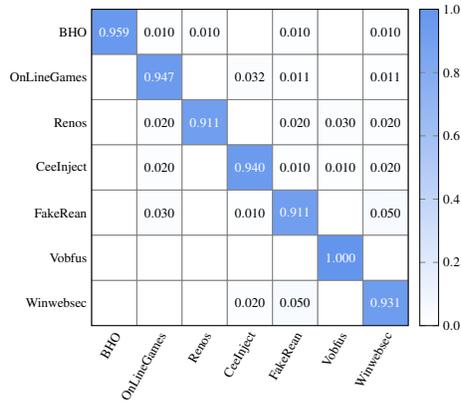

\centering
\input figures/conf_w2v_dnn.tex
\caption{Confusion matrix for Word2Vec-CNN}
\label{fig:/word2vec_dnn_v31_w1_cm}
\end{figure}

From Figure~\ref{fig:/word2vec_dnn_all}, we see that
our best accuracy achieved using a Word2Vec-CNN 
architecture is~94\%.
Figure~\ref{fig:/word2vec_dnn_v31_w1_cm} gives the confusion matrix 
for this best Word2Vec-CNN model. We see that the Fakerean family is
relatively often misclassified as OnLineGames or Winwebsec.
In our previous experiments, we have observed that Fakerean is 
generally the most challenging family to correctly classify. 

\subsection{Robustness Experiments}

Since malware often evolves, it is
advantageous if a classification technique
is robust to change in malware samples. Therefore,
as a final set of experiments, we explore the robustness of the various models 
considered in the previous sections.

For our robustness experiments, we consider binary 
classification, using the Winwebsec and Fakerean families.
And, as with our previous experiments, we
only consider the~31 most frequent opcodes.
We scramble the opcode sequences by a factor of~10\%, 20\%, 30\%, and~40\%.
For example, a scrambling factor of~10\%\ means that we select a block
that contains~10\%\ of the opcodes in a sample and we randomly shuffle
the opcodes in this selected block.
The resulting scrambled opcode sequence is then used 
to generate HMM2Vec and Word2Vec feature vectors.


\begin{figure}[!htb]
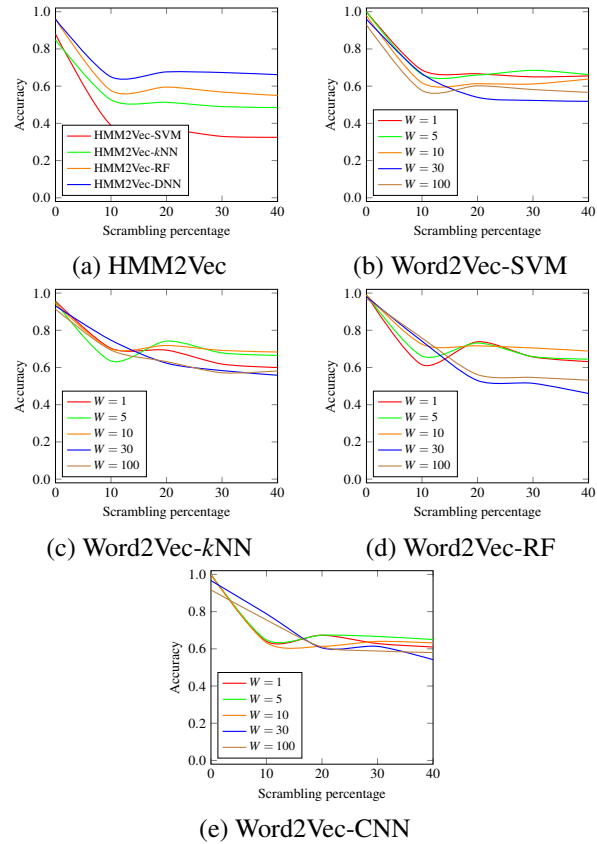

\centering
\begin{tabular}{cc}
\input figures/robust_hmm.tex & \input figures/robust_w2v.tex \\
(a) HMM2Vec & (b) Word2Vec-SVM \\
\input figures/robust_w2v_knn.tex & \input figures/robust_w2v_rf.tex \\
(c) Word2Vec-$k$NN & (d) Word2Vec-RF \\
\multicolumn{2}{c}{\input figures/robust_w2v_dnn.tex} \\
\multicolumn{2}{c}{(e) Word2Vec-CNN}
\end{tabular}
\caption{Robustness for Winwebsec and Fakerean binary classification}
\label{fig:robust}
\end{figure}

The resulting HMM2Vec experiments are summarized in Figure~\ref{fig:robust}~(a).
For the Word2Vec experiments, we test various windows sizes~$W$ in each case.
These Word2Vec
experiments are summarized in Figures~\ref{fig:robust}~(b) through~(e).

In general, Figure~\ref{fig:robust} indicates that
our opcode feature embedding techniques are not
particularly robust with respect to the shuffling experiments
that we have performed. In every case, the accuracy drops to~80\%\ or
less with just~10\%\ scrambling. Yet, there are some noteworthy results. For example,
we see that SVM far outperforms the other classifiers on the HMM2Vec features.
In contrast, for the Word2Vec features, small window sizes perform
best and, with respect to the classifiers, 
SVM performs relatively well, although several other classifiers are competitive. 
Another interesting observation
is that there is not much decline in any of the techniques after 
about~10\%\ scrambling.

\section{Conclusion and Future Work}\label{chap:conclusion}

In this paper, we considered engineered features for malware classification.
These engineered features were derived from opcode sequences
via a technique we refer to as HMM2Vec, and a parallel set of experiments
was conducted using
Word2Vec embeddings. We experimented with a 
diverse set of seven malware families. We also used four different classifiers
with each of the two engineered feature sets,
and we conducted a significant number of experiments to tune the
various hyperparameters of the machine learning algorithms.

\begin{figure}[!htb]
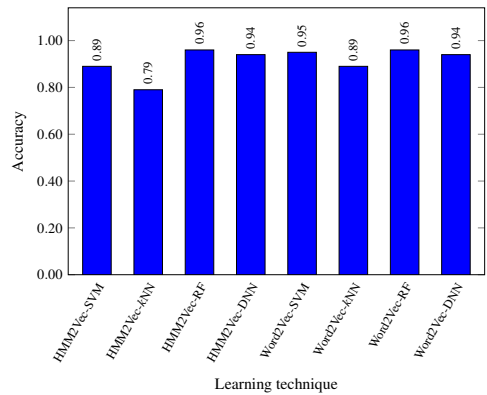

\centering
\input figures/best.tex
\caption{Best accuracies for HMM2Vec hybrid machine learning}\label{fig:/overall}
\end{figure}

Figure~\ref{fig:/overall} summarizes the best accuracies for our Word2Vec 
and HMM2Vec hybrid classification techniques. From Figure~\ref{fig:/overall} we see that
that HMM2Vec-RF and Word2Vec-RF attained the best results, 
with 96\%\ accuracy when classifying a balanced set of samples from seven families.
All of the hybrid machine learning techniques 
based on Word2Vec embeddings performed well, while the HMM2Vec
results were more mixed. This may be due to the relatively limited 
number of options considered when training
HMMs in our experiments, as compared to Word2Vec.

Almost all our hybrid machine learning techniques classified 
samples from BHO, Vobfus, and Renos with very high accuracy. 
We observed that the Winwebsec and OnLineGames samples
were often misclassified as Fakerean. The percentage of these 
misclassification was higher in HMM2Vec than Word2Vec, 
and accounts for most of the difference between these classification
techniques.

We also performed experiments that showed that 
none of the techniques considered in this paper is particularly
robust with respect to feature scrambling.
The Word2Vec models were more consistent in these
experiments, but the best HMM2Vec models performed
as well as the best Word2Vec models.

A major advantage of Word2Vec was its faster training time. We found that 
generating HMM2Vec features was 
slower than generating Word2Vec features
by a factor of about~15.
This vast difference between the two
cases was primarily due to the need to train multiple HMMs (i.e., multiple random restarts)
in cases where the amount of training data is relatively small.
Word2Vec can be trained on short opcode sequences,
since a larger window size~$W$ effectively inflates
the number of training samples that are available.

As future extension of this research, similar experiments could be 
performed on a larger and more diverse set of malware families. 
In this research, we only considered opcode sequences---analogous
experiments on other features, such as byte~$n$-grams or
dynamic features such as API
calls would be interesting.

Further experiments involving the many parameters found in the various 
machine learning techniques considered here would be worthwhile.
To mention just one of many such examples, 
additional combinations of window sizes and feature vector
lengths could be considered in Word2Vec.
Finally, other machine learning paradigms would be worth considering
in the context of malware detection based on vector embedding features.
Examples of other machine learning approaches that could be
advantageous for this problem include adversarial networks 
and reinforcement learning.

\bibliographystyle{apalike}

{\small
\bibliography{references.bib}
}

\end{document}

%% file: preamble.tex
\usepackage[scr=boondoxo,scrscaled=1.05]{mathalfa}
\usepackage[pdftex]{graphicx}
\usepackage{amsmath,amssymb,hhline,tikz,mathtools}
\usepackage{enumerate}
\usetikzlibrary{shapes.geometric,decorations.markings}
\usetikzlibrary{calc}
\usepackage[title]{appendix}
\usepackage{etoolbox}
\patchcmd{\appendices}{\quad}{: }{}{}
\usepackage{algorithm}
\usepackage{algorithmicx}
\usepackage{algpseudocode}
\usepackage{rotating}
\usepackage{afterpage}
\usepackage{xstring}
\usepackage{xcolor}
\usepackage{cmap}
\usepackage{mathdots}
\usepackage{colortbl}
\usepackage{booktabs}
\usepackage{textcomp}     
\usetikzlibrary{backgrounds}
\usepackage{pgfplots}
\pgfplotsset{compat=1.12}
\pgfmathsetseed{5112}
\usepackage{multirow}
\usepackage{pgfplotstable}
\usepackage{adjustbox}
\usepackage{diagbox}

\PassOptionsToPackage{hyphens}{url}
\usepackage{hyperref}
\hypersetup{
    pdftitle={Word Embedding Techniques for Malware Classification},
    pdfauthor={Mark Stamp},
    bookmarksnumbered=true,     
    bookmarksopen=true,         
    bookmarksopenlevel=3,       
    colorlinks=true,linkcolor=black,citecolor=black,urlcolor=blue,filecolor=blue,
    pdfstartview=Fit,           
    pdfpagemode=UseOutlines,
    pdfpagelayout=SinglePage
}

\definecolor{darkgreen}{rgb}{0.125,0.5,0.169}

\algrenewcommand{\algorithmiccomment}[1]{{\color{red}{\tt //}\ #1}}
\algnewcommand{\Initialize}[1]{%
  \State \textbf{Initialize:}
  \Statex \hspace*{\algorithmicindent}\parbox[t]{.8\linewidth}{\raggedright #1}
}
\algnewcommand{\Given}[1]{%
  \State \textbf{Given:}
  \Statex \hspace*{\algorithmicindent}\parbox[t]{.8\linewidth}{\raggedright #1}
}

\long\def\symbolfootnotetext[#1]#2{\begingroup%
\def\thefootnote{\fnsymbol{footnote}}\footnotetext[#1]{#2}\endgroup}

\let\oldsqrt\sqrt
\def\sqrt{\mathpalette\DHLhksqrt}
\def\DHLhksqrt#1#2{%
\setbox0=\hbox{$#1\oldsqrt{#2\,}$}\dimen0=\ht0
\advance\dimen0-0.2\ht0
\setbox2=\hbox{\vrule height\ht0 depth -\dimen0}%
{\box0\lower0.4pt\box2}}

\def\clap#1{\hbox to 0pt{\hss#1\hss}}

\allowdisplaybreaks

\def\figureFastSpeed{s}\def\figureSpeed{f}
\let\figureFastSpeed=\figureSpeed
\def\selectFigureSpeed#1#2{
\if\figureSpeed\figureFastSpeed #1\else #2\fi}

\def\srowvecc#1#2{(\!\begin{array}{cc} 
      \noexpandarg\IfBeginWith{#1}{-}{\! #1}{#1}
    & #2\kern-0.5pt\end{array}\!)}
\def\rowvecc#1#2{\left(\!\begin{array}{cc} 
      \noexpandarg\IfBeginWith{#1}{-}{\! #1}{#1}
    & #2\kern-0.5pt\end{array}\!\right)}
\def\rowveccc#1#2#3{\left(\!\begin{array}{ccc} 
      \noexpandarg\IfBeginWith{#1}{-}{\! #1}{#1}
    & #2 
    & #3\kern-0.5pt\end{array}\!\right)}
\def\rowvecccc#1#2#3#4{\left(\!\begin{array}{cccc}
      \noexpandarg\IfBeginWith{#1}{-}{\! #1}{#1}
    & #2 
    & #3 
    & #4\kern-0.5pt\end{array}\!\right)}
\def\srowvecccc#1#2#3#4{\bigl(\!\begin{array}{cccc}
      \noexpandarg\IfBeginWith{#1}{-}{\! #1}{#1}
    & #2 
    & #3 
    & #4\kern-0.5pt\end{array}\!\bigr)}
\def\rowveccccc#1#2#3#4#5{\left(\!\begin{array}{ccccc} 
      \noexpandarg\IfBeginWith{#1}{-}{\! #1}{#1}
    & #2
    & #3
    & #4
    & #5\kern-0.5pt\end{array}\!\right)}
\def\srowvecccccc#1#2#3#4#5#6{(\!\begin{array}{cccccc} 
      \noexpandarg\IfBeginWith{#1}{-}{\! #1}{#1}
    & #2
    & #3
    & #4
    & #5
    & #6\kern-0.5pt\end{array}\!)}
\def\rowvecccccc#1#2#3#4#5#6{\left(\!\begin{array}{cccccc} 
      \noexpandarg\IfBeginWith{#1}{-}{\! #1}{#1}
    & #2
    & #3
    & #4
    & #5
    & #6\kern-0.5pt\end{array}\!\right)}

\def\figureType{*}\def\figureSlowType{slowType}
\def\selectFigureType#1#2{
\if\figureType\figureSlowType #1\else #2\fi}
\makeatletter
\newcommand{\lowsub}[1]{\mathpalette{\raisem@th{#1}}}
\newcommand{\raisem@th}[3]{\raisebox{-#1}{$#2#3$}}
\makeatother
\def\halfthin{\kern 0.083em}

\pgfkeys{
    /pgf/number format/precision=2, 
    /pgf/number format/fixed zerofill=true }
    
\pgfplotstableset{
    /color cells/min/.initial=0,
    /color cells/max/.initial=1000,
    /color cells/textcolor/.initial=,
    color cells/.code={%
        \pgfqkeys{/color cells}{#1}%
        \pgfkeysalso{%
            postproc cell content/.code={%
                \begingroup
                \pgfkeysgetvalue{/pgfplots/table/@preprocessed cell content}\value
\ifx\value\empty
\endgroup
\else
                \pgfmathfloatparsenumber{\value}%
                \pgfmathfloattofixed{\pgfmathresult}%
                \let\value=\pgfmathresult
                \pgfplotscolormapaccess
                    [\pgfkeysvalueof{/color cells/min}:\pgfkeysvalueof{/color cells/max}]%
                    {\value}%
                    {\pgfkeysvalueof{/pgfplots/colormap name}}%
                \pgfkeysgetvalue{/pgfplots/table/@cell content}\typesetvalue
                \pgfkeysgetvalue{/color cells/textcolor}\textcolorvalue
                \toks0=\expandafter{\typesetvalue}%
                \xdef\temp{%
                    \noexpand\pgfkeysalso{%
                        @cell content={%
                            \noexpand\cellcolor[rgb]{\pgfmathresult}%
                            \noexpand\definecolor{mapped color}{rgb}{\pgfmathresult}%
                            \ifx\textcolorvalue\empty
                            \else
                                \noexpand\color{\textcolorvalue}%
                            \fi
                            \the\toks0 %
                        }%
                    }%
                }%
                \endgroup
                \temp
\fi
            }%
        }%
    }
}

\makeatletter
\newcommand*\bigcdot{\mathpalette\bigcdot@{.5}}
\newcommand*\bigcdot@[2]{\mathbin{\vcenter{\hbox{\scalebox{#2}{$\m@th#1\bullet$}}}}}
\makeatother

\def\k{\kern 2.75pt}

\newlength{\xxxxx}
\def\z{\phantom{0}}

%% file: figures/samples.tex
\begin{tikzpicture}[scale=0.475, every node/.style={scale=1.0}]
    \begin{axis}[
        width  = 1.0*\textwidth,
        height = 8cm,
        ymin=0,ymax=2850,
        ytick={0,500,1000,1500,2000,2500},
        major x tick style = transparent,
        ybar=5*\pgflinewidth,
        bar width=5.75pt,
        xlabel = {Family name},
        ylabel = {Number of samples},
        xlabel shift = -8pt,
        ylabel shift = -2pt,
        symbolic x coords={
VBInject,
Winwebsec,
Renos,
OnLineGames,
BHO,
Startpage,
Adload,
Vobfus,
CeeInject,
Lolyda.BF,
Zbot,
FakeRean,
Wintrim.BX,
Agent,
Allaple.A,
VB,
Cycbot.G,
Vundo,
Toga!rfn,
Rimecud.A,
Hotbar,
Obfuscator,
Injector,
Small,
Bulta!rfn,
Bifrose,
Zegost,
Drstwex.A,
Alureon,
DelfInject,
Zwangi,
Sinowal.gen!Y,
Hupigon,
FakeSysdef,
Koutodoor.E,
Bancos,
Banload
	},
        xtick={
VBInject,
Winwebsec,
Renos,
OnLineGames,
BHO,
Startpage,
Adload,
Vobfus,
CeeInject,
Lolyda.BF,
Zbot,
FakeRean,
Wintrim.BX,
Agent,
Allaple.A,
VB,
Cycbot.G,
Vundo,
Toga!rfn,
Rimecud.A,
Hotbar,
Obfuscator,
Injector,
Small,
Bulta!rfn,
Bifrose,
Zegost,
Drstwex.A,
Alureon,
DelfInject,
Zwangi,
Sinowal.gen!Y,
Hupigon,
FakeSysdef,
Koutodoor.E,
Bancos,
Banload
	},
	y tick label style={
		font=\small,
    		/pgf/number format/.cd,
   		fixed,
   		fixed zerofill,
		1000 sep={},
    		precision=0},
        x tick label style={
        		rotate=60,
		font=\footnotesize\tt,
		anchor=north east,
		inner sep=0mm
		},
        nodes near coords,
        every node near coord/.append style={rotate=90, 
        								   anchor=west,
								   font=\footnotesize,
								   /pgf/number format/.cd,
								   	fixed zerofill,
									1000 sep={},
									precision=0
								   },
        enlarge x limits=0.02,
    ]
\addplot[fill=blue,opacity=1.00] 
coordinates {
(VBInject,2431)
(Winwebsec,2302)
(Renos,1567)
(OnLineGames,1508)
(BHO,1396)
(Startpage,1284)
(Adload,1219)
(Vobfus,1107)
(CeeInject,1077)
(Lolyda.BF,1076)
(Zbot,1020)
(FakeRean,1017)
(Wintrim.BX,1000)
(Agent,988)
(Allaple.A,986)
(VB,978)
(Cycbot.G,858)
(Vundo,853)
(Toga!rfn,736)
(Rimecud.A,705)
(Hotbar,688)
(Obfuscator,681)
(Injector,675)
(Small,645)
(Bulta!rfn,569)
(Bifrose,546)
(Zegost,544)
(Drstwex.A,542)
(Alureon,539)
(DelfInject,504)
(Zwangi,460)
(Sinowal.gen!Y,427)
(Hupigon,426)
(FakeSysdef,398)
(Koutodoor.E,375)
(Bancos,361)
(Banload,311)
};
\end{axis}
\end{tikzpicture}

%% file: figures/opcodes.tex
\begin{tikzpicture}[scale=0.7, every node/.style={scale=1.0},rotate=-90]
    \begin{axis}[
        width  = 0.7*\textwidth,
        height = 10cm,
        ymin=0.0,ymax=20.0,
        ytick={0,5,10,15,20},
        major x tick style = transparent,
        ybar=5*\pgflinewidth,
        bar width=5.0pt,
        xlabel = {Opcode},
        xlabel style={rotate=180},
        ylabel = {Percentage of total},
        yticklabel pos=right,
        symbolic x coords={
        		mov,
		add,
		push,
		pop,
		inc,
		call,
		xor,
		dec,
		cmp,
		xchg,
		adc,
		sub,
		or,
		and,
		sbb,
		test,
		lea,
		jmp,
		je,
		ret,
		in,
		out,
		jne,
		imul,
		int,
		scas,
		nop,
		lods,
		int3,
		stos,
		pushl
	},
        xtick={
        		mov,
		add,
		push,
		pop,
		inc,
		call,
		xor,
		dec,
		cmp,
		xchg,
		adc,
		sub,
		or,
		and,
		sbb,
		test,
		lea,
		jmp,
		je,
		ret,
		in,
		out,
		jne,
		imul,
		int,
		scas,
		nop,
		lods,
		int3,
		stos,
		pushl
	},
	y tick label style={
		rotate=90,
    		/pgf/number format/.cd,
   		fixed,
   		fixed zerofill,
    		precision=0},
        x tick label style={
        		rotate=90,
		font=\small\tt,
		},
        nodes near coords,
        every node near coord/.append style={rotate=90, 
        								   anchor=west,
								   font=\footnotesize,
								   /pgf/number format/.cd,
								   	fixed zerofill,
									precision=2
								   },
        enlarge x limits=0.03,
    ]
\addplot[fill=blue,opacity=1.00] 
coordinates {
(mov,15.98601457)
(add,8.894469007)
(push,8.688018107)
(pop,5.116886966)
(inc,4.036563331)
(call,3.58227785)
(xor,3.247421036)
(dec,2.912125078)
(cmp,2.449324562)
(xchg,2.327350132)
(adc,2.209308681)
(sub,2.193354595)
(or,2.114786541)
(and,2.072056988)
(sbb,1.863930756)
(test,1.754845012)
(lea,1.635096766)
(jmp,1.583791363)
(je,1.371370569)
(ret,1.294958994)
(in,1.137655406)
(out,1.096099533)
(jne,0.935457063)
(imul,0.765446532)
(int,0.672325374)
(scas,0.609074562)
(nop,0.591594267)
(lods,0.566211761)
(int3,0.510860638)
(stos,0.501433664)
(pushl,0.488330162)
};
\end{axis}
\end{tikzpicture}

%% file: figures/w2v_svm_2.tex
\begin{tikzpicture}[scale=0.6, every node/.style={scale=1.0}]
    \begin{axis}[
        width  = 0.8*\textwidth,
        height = 8cm,
        ymin=0.0,ymax=1.25,
        ytick={0,0.2,0.4,0.6,0.8,1.0},
        major x tick style = transparent,
        ybar=5*\pgflinewidth,
        bar width=11.0pt,
        xlabel = {Window size},
        ylabel = {Accuracy},
        symbolic x coords={$W=1$,$W=5$,$W=10$,$W=30$,$W=100$},
        xtick=data,
	y tick label style={
    		/pgf/number format/.cd,
   		fixed,
   		fixed zerofill,
    		precision=2},
        x tick label style={
		font=\small,
		},
        nodes near coords,
        every node near coord/.append style={rotate=90, 
        								   anchor=west,
								   font=\footnotesize,
								   /pgf/number format/.cd,
								   	fixed zerofill,
									precision=4
								   },
        enlarge x limits=0.175,
        legend cell align=left,
        legend pos=south east,
    ]
\addplot[fill=red,opacity=1.00] 
coordinates {
($W=1$,0.9983)
($W=5$,0.9816)
($W=10$,0.9533)
($W=30$,0.9033)
($W=100$,0.8383)
};
\addplot[fill=blue,opacity=1.00] 
coordinates {
($W=1$,0.9966)
($W=5$,0.9966)
($W=10$,0.9850)
($W=30$,0.9400)
($W=100$,0.9333)
};
\addplot[fill=green,opacity=1.00] 
coordinates {
($W=1$,1.00)
($W=5$,0.9933)
($W=10$,0.9983)
($W=30$,0.9816)
($W=100$,0.9533)
};
\legend{20 opcodes,31 opcodes,40 opcodes}
\end{axis}
\end{tikzpicture}

%% file: figures/conf_hmm_svm.tex
\begin{tikzpicture}[scale=0.5]
    \begin{axis}[
        width=10cm,
        height=10cm,
	colormap={bluewhite}{color=(white) rgb255=(100,149,237)},
        xticklabels={BHO,OnLineGames,Renos,CeeInject,FakeRean,Vobfus,Winwebsec},
        xtick={0,...,6},
        xtick style={draw=none},
	xticklabel style={anchor=east,rotate=60,yshift=-5pt},
        yticklabels={BHO,OnLineGames,Renos,CeeInject,FakeRean,Vobfus,Winwebsec},
        ytick={0,...,6},
        ytick style={draw=none},
        enlargelimits=false,
        colorbar,
        colorbar style={
            ytick={0.0,0.2,0.4,0.6,0.8,1.0},
            yticklabels={0.0,0.2,0.4,0.6,0.8,1.0},
            yticklabel={\pgfmathprintnumber\tick},
            yticklabel style={
            		/pgf/number format/fixed,
			/pgf/number format/precision=1}
        },
        point meta min=0.0,
        point meta max=1.0,
        nodes near coords={\pgfmathprintnumber\pgfplotspointmeta},
        nodes near coords black white/.style={
            small value/.style={
                yshift=-7pt,
                text=black,
                /pgf/number format/fixed,
                /pgf/number format/precision=3
            },
            large value/.style={
                yshift=-7pt,
                text=white,
                /pgf/number format/fixed,
                /pgf/number format/precision=3
            },
            every node near coord/.style={
                check for zero/.code={
                    \pgfmathfloatifflags{\pgfplotspointmeta}{0}{
                        \pgfkeys{/tikz/coordinate}
                    }{
                        \begingroup
                        \pgfkeys{/pgf/fpu}
                        \pgfmathparse{\pgfplotspointmeta<#1}
                        \global\let\result=\pgfmathresult
                        \endgroup
                        %
                        %
                        \pgfmathfloatcreate{1}{1.0}{0}
                        \let\ONE=\pgfmathresult
                        \ifx\result\ONE
                            \pgfkeysalso{/pgfplots/small value}
                        \else
                            \pgfkeysalso{/pgfplots/large value}
                        \fi
                    }
                },
                check for zero,
            },
        },
        nodes near coords black white=0.5,
    ]
        \addplot[
            matrix plot,
            mesh/cols=7,
            point meta=explicit,draw=gray
        ] table [meta=C] {
            x y C
 0  0 0.942
 1  0 0.010
 2  0 0.003
 3  0 0.014
 4  0 0.031
 5  0 0.000
 6  0 0.000
 0  1 0.019
 1  1 0.856
 2  1 0.003
 3  1 0.013
 4  1 0.077
 5  1 0.000
 6  1 0.032
 0  2 0.000
 1  2 0.023
 2  2 0.912
 3  2 0.030
 4  2 0.039
 5  2 0.003
 6  2 0.020
 0  3 0.000
 1  3 0.030
 2  3 0.017
 3  3 0.886
 4  3 0.044
 5  3 0.000
 6  3 0.023
 0  4 0.010
 1  4 0.040
 2  4 0.017
 3  4 0.040
 4  4 0.845
 5  4 0.000
 6  4 0.047
 0  5 0.000
 1  5 0.009
 2  5 0.009
 3  5 0.006
 4  5 0.006
 5  5 0.966
 6  5 0.003
 0  6 0.000
 1  6 0.040
 2  6 0.036
 3  6 0.036
 4  6 0.091
 5  6 0.000
 6  6 0.796
        };
    \end{axis}
\end{tikzpicture}
%

%% file: figures/hmm_knn2.tex
\begin{tikzpicture}[scale=0.6]
\begin{axis}[smooth,
		   width=0.7\textwidth,
		   height=0.575\textwidth,
	 	   x tick label style={
   		 	/pgf/number format/.cd,
			/pgf/number format/1000 sep={},
   			fixed,
   			fixed zerofill,
    			precision=0
		   },
	 	   y tick label style={
    		 	/pgf/number format/.cd,
   			fixed,
   			fixed zerofill,
    			precision=2
		    },
                    xmin=0,xmax=101,
                    ymin=0.74,ymax=0.96,
                    xtick={0,10,20,30,40,50,60,70,80,90,100},
                    ytick={0.75,0.80,0.85,0.90,0.95},
                    xlabel={Number of neighbors~$k$},
                    ylabel={Accuracy}] 
\addplot[color=blue,ultra thick,
no marks] coordinates {
(1,0.935428571)
(2,0.927)
(3,0.924)
(4,0.921428571)
(5,0.918714286)
(6,0.914285714)
(7,0.911285714)
(8,0.908571429)
(9,0.903714286)
(10,0.899857143)
(11,0.896571429)
(12,0.894714286)
(13,0.893428571)
(14,0.890714286)
(15,0.888428571)
(16,0.885571429)
(17,0.883285714)
(18,0.880142857)
(19,0.878571429)
(20,0.874428571)
(21,0.872714286)
(22,0.869857143)
(23,0.867571429)
(24,0.865285714)
(25,0.863428571)
(26,0.859857143)
(27,0.858428571)
(28,0.855428571)
(29,0.854857143)
(30,0.853428571)
(31,0.851285714)
(32,0.848571429)
(33,0.845)
(34,0.842857143)
(35,0.840285714)
(36,0.838571429)
(37,0.836571429)
(38,0.835714286)
(39,0.833428571)
(40,0.831714286)
(41,0.830714286)
(42,0.829285714)
(43,0.827571429)
(44,0.824857143)
(45,0.822857143)
(46,0.820714286)
(47,0.820142857)
(48,0.816285714)
(49,0.816142857)
(50,0.814428571)
(51,0.812)
(52,0.811285714)
(53,0.809571429)
(54,0.808)
(55,0.806714286)
(56,0.806)
(57,0.805)
(58,0.804428571)
(59,0.802857143)
(60,0.800857143)
(61,0.800571429)
(62,0.798857143)
(63,0.796571429)
(64,0.795428571)
(65,0.794857143)
(66,0.794)
(67,0.792571429)
(68,0.790857143)
(69,0.790428571)
(70,0.789714286)
(71,0.789571429)
(72,0.788428571)
(73,0.788)
(74,0.787428571)
(75,0.786714286)
(76,0.785428571)
(77,0.783714286)
(78,0.781142857)
(79,0.780857143)
(80,0.778571429)
(81,0.777285714)
(82,0.775142857)
(83,0.775428571)
(84,0.774714286)
(85,0.773857143)
(86,0.772571429)
(87,0.771)
(88,0.770571429)
(89,0.770285714)
(90,0.769)
(91,0.767142857)
(92,0.764571429)
(93,0.763142857)
(94,0.762714286)
(95,0.762)
(96,0.762)
(97,0.759428571)
(98,0.757428571)
(99,0.756714286)
(100,0.756714286)
};
\end{axis}
\end{tikzpicture}

%% file: figures/conf_hmm_rf_grid.tex
\begin{tikzpicture}[scale=0.5]
    \begin{axis}[
        width=10cm,
        height=10cm,
	colormap={bluewhite}{color=(white) rgb255=(100,149,237)},
        xticklabels={BHO,OnLineGames,Renos,CeeInject,FakeRean,Vobfus,Winwebsec},
        xtick={0,...,6},
        xtick style={draw=none},
	xticklabel style={anchor=east,rotate=60,yshift=-5pt},
        yticklabels={BHO,OnLineGames,Renos,CeeInject,FakeRean,Vobfus,Winwebsec},
        ytick={0,...,6},
        ytick style={draw=none},
        enlargelimits=false,
        colorbar,
        colorbar style={
            ytick={0.0,0.2,0.4,0.6,0.8,1.0},
            yticklabels={0.0,0.2,0.4,0.6,0.8,1.0},
            yticklabel={\pgfmathprintnumber\tick},
            yticklabel style={
            		/pgf/number format/fixed,
			/pgf/number format/precision=1}
        },
        point meta min=0.0,
        point meta max=1.0,
        nodes near coords={\pgfmathprintnumber\pgfplotspointmeta},
        nodes near coords black white/.style={
            small value/.style={
                yshift=-7pt,
                text=black,
                /pgf/number format/fixed,
                /pgf/number format/precision=3
            },
            large value/.style={
                yshift=-7pt,
                text=white,
                /pgf/number format/fixed,
                /pgf/number format/precision=3
            },
            every node near coord/.style={
                check for zero/.code={
                    \pgfmathfloatifflags{\pgfplotspointmeta}{0}{
                        \pgfkeys{/tikz/coordinate}
                    }{
                        \begingroup
                        \pgfkeys{/pgf/fpu}
                        \pgfmathparse{\pgfplotspointmeta<#1}
                        \global\let\result=\pgfmathresult
                        \endgroup
                        %
                        %
                        \pgfmathfloatcreate{1}{1.0}{0}
                        \let\ONE=\pgfmathresult
                        \ifx\result\ONE
                            \pgfkeysalso{/pgfplots/small value}
                        \else
                            \pgfkeysalso{/pgfplots/large value}
                        \fi
                    }
                },
                check for zero,
            },
        },
        nodes near coords black white=0.5,
    ]
        \addplot[
            matrix plot,
            mesh/cols=7,
            point meta=explicit,draw=gray
        ] table [meta=C] {
            x y C
 0  0 0.973
 1  0 0.007
 2  0 0.000
 3  0 0.007
 4  0 0.007
 5  0 0.003
 6  0 0.003
 0  1 0.000
 1  1 0.920
 2  1 0.006
 3  1 0.025
 4  1 0.048
 5  1 0.000
 6  1 0.000
 0  2 0.000
 1  2 0.009
 2  2 0.955
 3  2 0.006
 4  2 0.030
 5  2 0.000
 6  2 0.000
 0  3 0.000
 1  3 0.007
 2  3 0.007
 3  3 0.960
 4  3 0.013
 5  3 0.007
 6  3 0.007
 0  4 0.000
 1  4 0.018
 2  4 0.000
 3  4 0.007
 4  4 0.961
 5  4 0.000
 6  4 0.014
 0  5 0.000
 1  5 0.003
 2  5 0.003
 3  5 0.000
 4  5 0.003
 5  5 0.990
 6  5 0.000
 0  6 0.000
 1  6 0.003
 2  6 0.007
 3  6 0.003
 4  6 0.045
 5  6 0.003
 6  6 0.938
        };
    \end{axis}
\end{tikzpicture}
%

%% file: figures/w2v_svm_linear.tex
\begin{tikzpicture}[scale=0.6, every node/.style={scale=1.0}]
    \begin{axis}[
        width  = 0.8*\textwidth,
        height = 8cm,
        ymin=0.0,ymax=1.12,
        ytick={0,0.2,0.4,0.6,0.8,1.0},
        major x tick style = transparent,
        ybar=5*\pgflinewidth,
        bar width=11.0pt,
        xlabel = {Embedding vector length},
        ylabel = {Accuracy},
        ylabel shift = -2.5pt,
        symbolic x coords={$N=2$, $N=31$,$N=100$},
	y tick label style={
    		/pgf/number format/.cd,
   		fixed,
   		fixed zerofill,
    		precision=1},
        xtick={$N=2$,$N=31$,$N=100$},
        x tick label style={
		font=\small,
		},
        nodes near coords,
        every node near coord/.append style={rotate=90, 
        								   anchor=west,
								   font=\footnotesize,
								   /pgf/number format/.cd,
								   	fixed zerofill,
									precision=2
								   },
        enlarge x limits=0.25,
        legend cell align=left,
        legend pos=south east,
    ]
\addplot[fill=red,opacity=1.00] 
coordinates {
($N=2$,0.88)
($N=31$,0.95)
($N=100$,0.93)
};
\addplot[fill=blue,opacity=1.00] 
coordinates {
($N=2$,0.89)
($N=31$,0.94)
($N=100$,0.94)
};
\addplot[fill=green,opacity=1.00] 
coordinates {
($N=2$,0.87)
($N=31$,0.95)
($N=100$,0.93)
};
\addplot[fill=yellow,opacity=1.00] 
coordinates {
($N=2$,0.85)
($N=31$,0.93)
($N=100$,0.92)
};
\addplot[fill=brown,opacity=1.00] 
coordinates {
($N=2$,0.81)
($N=31$,0.90)
($N=100$,0.92)
};
\legend{$W=1$,$W=5$,$W=10$,$W=30$,$W=100$}
\end{axis}
\end{tikzpicture}

%% file: figures/w2v_svm_rbf.tex
\begin{tikzpicture}[scale=0.6, every node/.style={scale=1.0}]
    \begin{axis}[
        width  = 0.8*\textwidth,
        height = 8cm,
        ymin=0.0,ymax=1.12,
        ytick={0,0.2,0.4,0.6,0.8,1.0},
        major x tick style = transparent,
        ybar=5*\pgflinewidth,
        bar width=11.0pt,
        xlabel = {Embedding vector length},
        ylabel = {Accuracy},
        ylabel shift = -2.5pt,
        symbolic x coords={$N=2$, $N=31$,$N=100$},
	y tick label style={
    		/pgf/number format/.cd,
   		fixed,
   		fixed zerofill,
    		precision=1},
        xtick={$N=2$,$N=31$,$N=100$},
        x tick label style={
		font=\small,
		},
        nodes near coords,
        every node near coord/.append style={rotate=90, 
        								   anchor=west,
								   font=\footnotesize,
								   /pgf/number format/.cd,
								   	fixed zerofill,
									precision=2
								   },
        enlarge x limits=0.25,
        legend cell align=left,
        legend pos=south east,
    ]
\addplot[fill=red,opacity=1.00] 
coordinates {
($N=2$,0.90)
($N=31$,0.95)
($N=100$,0.93)
};
\addplot[fill=blue,opacity=1.00] 
coordinates {
($N=2$,0.92)
($N=31$,0.93)
($N=100$,0.94)
};
\addplot[fill=green,opacity=1.00] 
coordinates {
($N=2$,0.91)
($N=31$,0.95)
($N=100$,0.94)
};
\addplot[fill=yellow,opacity=1.00] 
coordinates {
($N=2$,0.92)
($N=31$,0.93)
($N=100$,0.92)
};
\addplot[fill=brown,opacity=1.00] 
coordinates {
($N=2$,0.91)
($N=31$,0.91)
($N=100$,0.92)
};
\legend{$W=1$,$W=5$,$W=10$,$W=30$,$W=100$}
\end{axis}
\end{tikzpicture}

%% file: figures/w2v_knn2.tex
\begin{tikzpicture}[scale=0.6]
\begin{axis}[smooth,
		   width=0.7\textwidth,
		   height=0.575\textwidth,
	 	   x tick label style={
   		 	/pgf/number format/.cd,
			/pgf/number format/1000 sep={},
   			fixed,
   			fixed zerofill,
    			precision=0
		   },
	 	   y tick label style={
    		 	/pgf/number format/.cd,
   			fixed,
   			fixed zerofill,
    			precision=2
		    },
                    xmin=0,xmax=101,
                    ymin=0.85,ymax=0.97,
                    xtick={0,10,20,30,40,50,60,70,80,90,100},
                    ytick={0.86,0.88,0.90,0.92,0.94,0.96},
                    xlabel={Number of neighbors~$k$},
                    ylabel={Accuracy}] 
\addplot[color=blue,ultra thick,
no marks] coordinates {
(1,0.955142857)
(2,0.950285714)
(3,0.945)
(4,0.945428571)
(5,0.943142857)
(6,0.942714286)
(7,0.939714286)
(8,0.937857143)
(9,0.935285714)
(10,0.933714286)
(11,0.931142857)
(12,0.93)
(13,0.928857143)
(14,0.927428571)
(15,0.926571429)
(16,0.925428571)
(17,0.925)
(18,0.923714286)
(19,0.922428571)
(20,0.922285714)
(21,0.921142857)
(22,0.920857143)
(23,0.919)
(24,0.918428571)
(25,0.917142857)
(26,0.916714286)
(27,0.916142857)
(28,0.916571429)
(29,0.916428571)
(30,0.916285714)
(31,0.915428571)
(32,0.914714286)
(33,0.913428571)
(34,0.912571429)
(35,0.910571429)
(36,0.910142857)
(37,0.908571429)
(38,0.908)
(39,0.906428571)
(40,0.906)
(41,0.905142857)
(42,0.903714286)
(43,0.902142857)
(44,0.901)
(45,0.900285714)
(46,0.899571429)
(47,0.898428571)
(48,0.897285714)
(49,0.896857143)
(50,0.896428571)
(51,0.895857143)
(52,0.895428571)
(53,0.895428571)
(54,0.894714286)
(55,0.893857143)
(56,0.892857143)
(57,0.892142857)
(58,0.891714286)
(59,0.890857143)
(60,0.890857143)
(61,0.890428571)
(62,0.889428571)
(63,0.887714286)
(64,0.887857143)
(65,0.886714286)
(66,0.886)
(67,0.886142857)
(68,0.886142857)
(69,0.884142857)
(70,0.884571429)
(71,0.884857143)
(72,0.884142857)
(73,0.883714286)
(74,0.882142857)
(75,0.881571429)
(76,0.880714286)
(77,0.879571429)
(78,0.879428571)
(79,0.879428571)
(80,0.879571429)
(81,0.878)
(82,0.878428571)
(83,0.878142857)
(84,0.877)
(85,0.877285714)
(86,0.877142857)
(87,0.877285714)
(88,0.877)
(89,0.876571429)
(90,0.876)
(91,0.875571429)
(92,0.875857143)
(93,0.875142857)
(94,0.875714286)
(95,0.875428571)
(96,0.875142857)
(97,0.875)
(98,0.874285714)
(99,0.874857143)
(100,0.873857143)
};
\end{axis}
\end{tikzpicture}

%% file: figures/conf_w2v_rf.tex
\begin{tikzpicture}[scale=0.5]
    \begin{axis}[
        width=10cm,
        height=10cm,
	colormap={bluewhite}{color=(white) rgb255=(100,149,237)},
        xticklabels={BHO,OnLineGames,Renos,CeeInject,FakeRean,Vobfus,Winwebsec},
        xtick={0,...,6},
        xtick style={draw=none},
	xticklabel style={anchor=east,rotate=60,yshift=-5pt},
        yticklabels={BHO,OnLineGames,Renos,CeeInject,FakeRean,Vobfus,Winwebsec},
        ytick={0,...,6},
        ytick style={draw=none},
        enlargelimits=false,
        colorbar,
        colorbar style={
            ytick={0.0,0.2,0.4,0.6,0.8,1.0},
            yticklabels={0.0,0.2,0.4,0.6,0.8,1.0},
            yticklabel={\pgfmathprintnumber\tick},
            yticklabel style={
            		/pgf/number format/fixed,
			/pgf/number format/precision=1}
        },
        point meta min=0.0,
        point meta max=1.0,
        nodes near coords={\pgfmathprintnumber\pgfplotspointmeta},
        nodes near coords black white/.style={
            small value/.style={
                yshift=-7pt,
                text=black,
                /pgf/number format/fixed,
                /pgf/number format/precision=3
            },
            large value/.style={
                yshift=-7pt,
                text=white,
                /pgf/number format/fixed,
                /pgf/number format/precision=3
            },
            every node near coord/.style={
                check for zero/.code={
                    \pgfmathfloatifflags{\pgfplotspointmeta}{0}{
                        \pgfkeys{/tikz/coordinate}
                    }{
                        \begingroup
                        \pgfkeys{/pgf/fpu}
                        \pgfmathparse{\pgfplotspointmeta<#1}
                        \global\let\result=\pgfmathresult
                        \endgroup
                        %
                        %
                        \pgfmathfloatcreate{1}{1.0}{0}
                        \let\ONE=\pgfmathresult
                        \ifx\result\ONE
                            \pgfkeysalso{/pgfplots/small value}
                        \else
                            \pgfkeysalso{/pgfplots/large value}
                        \fi
                    }
                },
                check for zero,
            },
        },
        nodes near coords black white=0.5,
    ]
        \addplot[
            matrix plot,
            mesh/cols=7,
            point meta=explicit,draw=gray
        ] table [meta=C] {
            x y C
 0  0 0.977
 1  0 0.000
 2  0 0.000
 3  0 0.003
 4  0 0.013
 5  0 0.003
 6  0 0.003
 0  1 0.003
 1  1 0.970
 2  1 0.000
 3  1 0.010
 4  1 0.010
 5  1 0.007
 6  1 0.000
 0  2 0.000
 1  2 0.010
 2  2 0.949
 3  2 0.010
 4  2 0.010
 5  2 0.006
 6  2 0.016
 0  3 0.003
 1  3 0.013
 2  3 0.007
 3  3 0.953
 4  3 0.013
 5  3 0.000
 6  3 0.010
 0  4 0.000
 1  4 0.017
 2  4 0.007
 3  4 0.003
 4  4 0.959
 5  4 0.000
 6  4 0.014
 0  5 0.003
 1  5 0.000
 2  5 0.003
 3  5 0.000
 4  5 0.007
 5  5 0.986
 6  5 0.000
 0  6 0.003
 1  6 0.007
 2  6 0.007
 3  6 0.007
 4  6 0.037
 5  6 0.000
 6  6 0.940
        };
    \end{axis}
\end{tikzpicture}
%

%% file: figures/w2v_dnn_accuracy_overfit.tex
\begin{tikzpicture}[scale=0.45]
\begin{axis}[smooth,
		   width=0.7\textwidth,
		   height=0.575\textwidth,
	 	   x tick label style={
   		 	/pgf/number format/.cd,
			/pgf/number format/1000 sep={},
   			fixed,
   			fixed zerofill,
    			precision=0
		   },
	 	   y tick label style={
    		 	/pgf/number format/.cd,
   			fixed,
   			fixed zerofill,
    			precision=2
		    },
                    xmin=0,xmax=101,
                    ymin=0.685,ymax=1.015,
                    legend pos=south west,
                    xtick={0,20,40,60,80,100},
                    ytick={0.70,0.75,0.80,0.85,0.9,0.95,1.00},
                    xlabel={Epoch},
                    ylabel={Accuracy}] 
\addplot[color=blue,thick,
no marks] coordinates {
(1,0.8345)
(2,0.9114)
(3,0.9262)
(4,0.9359)
(5,0.9407)
(6,0.9427)
(7,0.9479)
(8,0.9488)
(9,0.9514)
(10,0.952)
(11,0.9573)
(12,0.958)
(13,0.9625)
(14,0.9611)
(15,0.967)
(16,0.9671)
(17,0.9682)
(18,0.9652)
(19,0.9688)
(20,0.9671)
(21,0.9712)
(22,0.9725)
(23,0.9736)
(24,0.9755)
(25,0.9743)
(26,0.9761)
(27,0.9775)
(28,0.9775)
(29,0.9775)
(30,0.9743)
(31,0.9755)
(32,0.9802)
(33,0.9804)
(34,0.978)
(35,0.9832)
(36,0.9814)
(37,0.9777)
(38,0.9793)
(39,0.9805)
(40,0.9827)
(41,0.9846)
(42,0.9839)
(43,0.9859)
(44,0.9857)
(45,0.9839)
(46,0.9834)
(47,0.9821)
(48,0.988)
(49,0.9866)
(50,0.9875)
(51,0.9875)
(52,0.9886)
(53,0.9884)
(54,0.9891)
(55,0.9902)
(56,0.9821)
(57,0.9873)
(58,0.9911)
(59,0.9907)
(60,0.9902)
(61,0.9912)
(62,0.9912)
(63,0.9896)
(64,0.9871)
(65,0.9875)
(66,0.9904)
(67,0.9911)
(68,0.9895)
(69,0.9923)
(70,0.9929)
(71,0.9937)
(72,0.9927)
(73,0.993)
(74,0.9934)
(75,0.9912)
(76,0.9961)
(77,0.9937)
(78,0.9921)
(79,0.9932)
(80,0.9946)
(81,0.9909)
(82,0.9816)
(83,0.9871)
(84,0.9921)
(85,0.9962)
(86,0.9964)
(87,0.9973)
(88,0.9975)
(89,0.9973)
(90,0.9945)
(91,0.997)
(92,0.9961)
(93,0.9966)
(94,0.9971)
(95,0.9962)
(96,0.9939)
(97,0.9934)
(98,0.9907)
(99,0.9855)
(100,0.9898)
};
\addplot[color=red,thick,
no marks] coordinates {
(1,0.9057)
(2,0.9086)
(3,0.9314)
(4,0.9343)
(5,0.9343)
(6,0.9386)
(7,0.94)
(8,0.9371)
(9,0.9443)
(10,0.9414)
(11,0.9486)
(12,0.9514)
(13,0.95)
(14,0.9371)
(15,0.9557)
(16,0.9543)
(17,0.95)
(18,0.9543)
(19,0.96)
(20,0.96)
(21,0.9586)
(22,0.9657)
(23,0.9671)
(24,0.9657)
(25,0.97)
(26,0.9629)
(27,0.9643)
(28,0.9614)
(29,0.9571)
(30,0.9657)
(31,0.9629)
(32,0.96)
(33,0.9643)
(34,0.9629)
(35,0.96)
(36,0.9643)
(37,0.9586)
(38,0.9629)
(39,0.9614)
(40,0.9629)
(41,0.9629)
(42,0.9586)
(43,0.9643)
(44,0.9671)
(45,0.9629)
(46,0.9671)
(47,0.9671)
(48,0.9671)
(49,0.9686)
(50,0.97)
(51,0.9643)
(52,0.97)
(53,0.9657)
(54,0.9671)
(55,0.9571)
(56,0.9671)
(57,0.9671)
(58,0.9714)
(59,0.9714)
(60,0.9729)
(61,0.9743)
(62,0.9786)
(63,0.9686)
(64,0.9686)
(65,0.9686)
(66,0.9757)
(67,0.9743)
(68,0.9714)
(69,0.9686)
(70,0.9729)
(71,0.9729)
(72,0.9686)
(73,0.9757)
(74,0.9757)
(75,0.9714)
(76,0.9729)
(77,0.9729)
(78,0.9714)
(79,0.9671)
(80,0.9714)
(81,0.97)
(82,0.9657)
(83,0.9686)
(84,0.97)
(85,0.97)
(86,0.97)
(87,0.9743)
(88,0.9729)
(89,0.9714)
(90,0.9643)
(91,0.9714)
(92,0.9729)
(93,0.9729)
(94,0.9714)
(95,0.9671)
(96,0.97)
(97,0.9643)
(98,0.97)
(99,0.9657)
(100,0.9643)
};
\legend{Train, Validation}
\end{axis}
\end{tikzpicture}

%% file: figures/w2v_dnn_loss_overfit.tex
\begin{tikzpicture}[scale=0.45]
\begin{axis}[smooth,
		   width=0.7\textwidth,
		   height=0.575\textwidth,
	 	   x tick label style={
   		 	/pgf/number format/.cd,
			/pgf/number format/1000 sep={},
   			fixed,
   			fixed zerofill,
    			precision=0
		   },
	 	   y tick label style={
    		 	/pgf/number format/.cd,
   			fixed,
   			fixed zerofill,
    			precision=1
		    },
                    xmin=0,xmax=101,
                    ymin=-0.02,ymax=0.62,
                    legend pos=north west,
                    xtick={0,20,40,60,80,100},
                    ytick={0.0,0.1,0.2,0.3,0.4,0.5,0.6},
                    xlabel={Epoch},
                    ylabel={Loss}] 
\addplot[color=blue,thick,
no marks] coordinates {
(1,0.5238)
(2,0.3191)
(3,0.2566)
(4,0.2198)
(5,0.2009)
(6,0.1896)
(7,0.1695)
(8,0.1623)
(9,0.1509)
(10,0.1471)
(11,0.1328)
(12,0.1267)
(13,0.1187)
(14,0.121)
(15,0.1093)
(16,0.1072)
(17,0.107)
(18,0.1051)
(19,0.1037)
(20,0.109)
(21,0.0944)
(22,0.0931)
(23,0.0861)
(24,0.0801)
(25,0.0804)
(26,0.078)
(27,0.0711)
(28,0.0801)
(29,0.0795)
(30,0.086)
(31,0.0808)
(32,0.0719)
(33,0.0667)
(34,0.0717)
(35,0.0599)
(36,0.0709)
(37,0.0648)
(38,0.0639)
(39,0.0612)
(40,0.0567)
(41,0.0532)
(42,0.0545)
(43,0.0491)
(44,0.0502)
(45,0.0648)
(46,0.0564)
(47,0.0585)
(48,0.0452)
(49,0.044)
(50,0.0423)
(51,0.0419)
(52,0.0434)
(53,0.0393)
(54,0.0361)
(55,0.0357)
(56,0.0592)
(57,0.0432)
(58,0.0325)
(59,0.033)
(60,0.0341)
(61,0.0319)
(62,0.0329)
(63,0.0348)
(64,0.043)
(65,0.0397)
(66,0.0322)
(67,0.0309)
(68,0.0311)
(69,0.0273)
(70,0.0244)
(71,0.0223)
(72,0.0245)
(73,0.023)
(74,0.0213)
(75,0.0267)
(76,0.0191)
(77,0.0223)
(78,0.0235)
(79,0.0206)
(80,0.0197)
(81,0.0375)
(82,0.0778)
(83,0.0454)
(84,0.027)
(85,0.0136)
(86,0.0139)
(87,0.0139)
(88,0.012)
(89,0.0116)
(90,0.0206)
(91,0.0141)
(92,0.015)
(93,0.0143)
(94,0.0134)
(95,0.0136)
(96,0.0195)
(97,0.021)
(98,0.0287)
(99,0.0521)
(100,0.0382)
};
\addplot[color=red,thick,
no marks] coordinates {
(1,0.322)
(2,0.3005)
(3,0.2557)
(4,0.221)
(5,0.2188)
(6,0.2259)
(7,0.2127)
(8,0.2073)
(9,0.2085)
(10,0.2008)
(11,0.1991)
(12,0.1909)
(13,0.1882)
(14,0.226)
(15,0.1786)
(16,0.1963)
(17,0.227)
(18,0.2048)
(19,0.2047)
(20,0.1819)
(21,0.1967)
(22,0.1942)
(23,0.1886)
(24,0.1829)
(25,0.1767)
(26,0.1976)
(27,0.2066)
(28,0.2094)
(29,0.2258)
(30,0.2056)
(31,0.1955)
(32,0.204)
(33,0.2245)
(34,0.2048)
(35,0.2158)
(36,0.1934)
(37,0.2277)
(38,0.2423)
(39,0.2501)
(40,0.231)
(41,0.2181)
(42,0.2656)
(43,0.2372)
(44,0.2506)
(45,0.2441)
(46,0.2623)
(47,0.2596)
(48,0.2295)
(49,0.2121)
(50,0.2371)
(51,0.2133)
(52,0.2277)
(53,0.2198)
(54,0.2538)
(55,0.2443)
(56,0.2722)
(57,0.2666)
(58,0.2533)
(59,0.2505)
(60,0.2419)
(61,0.2504)
(62,0.2421)
(63,0.236)
(64,0.256)
(65,0.2586)
(66,0.26)
(67,0.2536)
(68,0.2858)
(69,0.2647)
(70,0.2519)
(71,0.2637)
(72,0.2891)
(73,0.2604)
(74,0.2775)
(75,0.2998)
(76,0.3062)
(77,0.2707)
(78,0.3137)
(79,0.2958)
(80,0.3037)
(81,0.3324)
(82,0.3995)
(83,0.4236)
(84,0.3274)
(85,0.3349)
(86,0.3263)
(87,0.336)
(88,0.3436)
(89,0.3252)
(90,0.344)
(91,0.3313)
(92,0.3425)
(93,0.3417)
(94,0.3427)
(95,0.3676)
(96,0.3553)
(97,0.3566)
(98,0.3809)
(99,0.4192)
(100,0.3922)
};
\legend{Train, Validation}
\end{axis}
\end{tikzpicture}

%% file: figures/w2v_dnn_accuracy.tex
\begin{tikzpicture}[scale=0.45]
\begin{axis}[smooth,
		   width=0.7\textwidth,
		   height=0.575\textwidth,
	 	   x tick label style={
   		 	/pgf/number format/.cd,
			/pgf/number format/1000 sep={},
   			fixed,
   			fixed zerofill,
    			precision=0
		   },
	 	   y tick label style={
    		 	/pgf/number format/.cd,
   			fixed,
   			fixed zerofill,
    			precision=2
		    },
                    xmin=0,xmax=51,
                    ymin=0.685,ymax=0.965,
                    legend pos=south west,
                    xtick={0,10,20,30,40,50},
                    ytick={0.70,0.75,0.80,0.85,0.90,0.95},
                    xlabel={Epoch},
                    ylabel={Accuracy}] 
\addplot[color=blue,thick,
no marks] coordinates {
(1,0.6816071)
(2,0.80482143)
(3,0.8435714)
(4,0.86482143)
(5,0.8748214)
(6,0.89017856)
(7,0.89875)
(8,0.90660715)
(9,0.9080357)
(10,0.91)
(11,0.9114286)
(12,0.91732144)
(13,0.91785717)
(14,0.92035717)
(15,0.92464286)
(16,0.9269643)
(17,0.9276786)
(18,0.92732143)
(19,0.9333929)
(20,0.93035716)
(21,0.9351786)
(22,0.9332143)
(23,0.9382143)
(24,0.9392857)
(25,0.9383929)
(26,0.93892854)
(27,0.9426786)
(28,0.9430357)
(29,0.9417857)
(30,0.94125)
(31,0.94625)
(32,0.9475)
(33,0.94285715)
(34,0.9473214)
(35,0.9448214)
(36,0.94839287)
(37,0.94910717)
(38,0.94785714)
(39,0.94910717)
(40,0.94910717)
(41,0.95017856)
(42,0.9514286)
(43,0.9507143)
(44,0.95196426)
(45,0.95125)
(46,0.9525)
(47,0.9532143)
(48,0.9530357)
(49,0.95410717)
(50,0.95589286)
};
\addplot[color=red,thick,
no marks] coordinates {
(1,0.79285717)
(2,0.838571429)
(3,0.85857141)
(4,0.882857144)
(5,0.89142859)
(6,0.89142859)
(7,0.912857115)
(8,0.904285729)
(9,0.910000026)
(10,0.908571422)
(11,0.914285719)
(12,0.915714264)
(13,0.922857165)
(14,0.920000017)
(15,0.92428571)
(16,0.921428561)
(17,0.925714314)
(18,0.92428571)
(19,0.92428571)
(20,0.925714314)
(21,0.92428571)
(22,0.92428571)
(23,0.925714314)
(24,0.927142859)
(25,0.925714314)
(26,0.925714314)
(27,0.928571403)
(28,0.928571403)
(29,0.927142859)
(30,0.930000007)
(31,0.928571403)
(32,0.930000007)
(33,0.931428552)
(34,0.930000007)
(35,0.931428552)
(36,0.930000007)
(37,0.920000017)
(38,0.928571403)
(39,0.932857156)
(40,0.931428552)
(41,0.928571403)
(42,0.932857156)
(43,0.928571403)
(44,0.931428552)
(45,0.9342857)
(46,0.931428552)
(47,0.931428552)
(48,0.928571403)
(49,0.932857156)
(50,0.932857156)
};
\legend{Train, Validation}
\end{axis}
\end{tikzpicture}

%% file: figures/w2v_dnn_loss.tex
\begin{tikzpicture}[scale=0.45]
\begin{axis}[smooth,
		   width=0.7\textwidth,
		   height=0.575\textwidth,
	 	   x tick label style={
   		 	/pgf/number format/.cd,
			/pgf/number format/1000 sep={},
   			fixed,
   			fixed zerofill,
    			precision=0
		   },
	 	   y tick label style={
    		 	/pgf/number format/.cd,
   			fixed,
   			fixed zerofill,
    			precision=2
		    },
                    xmin=0,xmax=51,
                    ymin=0.05,ymax=1.05,
                    legend pos=north west,
                    xtick={0,10,20,30,40,50},
                    ytick={0.10,0.25,0.40,0.55,0.70,0.85,1.00},
                    xlabel={Epoch},
                    ylabel={Loss}] 
\addplot[color=blue,thick,
no marks] coordinates {
(1,0.993583475)
(2,0.623260944)
(3,0.526650244)
(4,0.462146021)
(5,0.432060893)
(6,0.399390475)
(7,0.375822915)
(8,0.357584706)
(9,0.341194986)
(10,0.329667254)
(11,0.315581441)
(12,0.302646858)
(13,0.292861869)
(14,0.28711817)
(15,0.27550378)
(16,0.264240122)
(17,0.260284946)
(18,0.25525384)
(19,0.244352488)
(20,0.242318082)
(21,0.234674936)
(22,0.233480839)
(23,0.228449663)
(24,0.220132955)
(25,0.216236781)
(26,0.212652539)
(27,0.208110942)
(28,0.204027626)
(29,0.204481677)
(30,0.200427046)
(31,0.198242775)
(32,0.192677604)
(33,0.197981449)
(34,0.185909744)
(35,0.189982233)
(36,0.181657344)
(37,0.181050217)
(38,0.178739839)
(39,0.178575353)
(40,0.173580302)
(41,0.175041577)
(42,0.170569214)
(43,0.173547392)
(44,0.165685829)
(45,0.167142163)
(46,0.16466791)
(47,0.161820625)
(48,0.159894938)
(49,0.158430053)
(50,0.157113887)
};
\addplot[color=red,thick,
no marks] coordinates {
(1,0.665807072)
(2,0.528993826)
(3,0.468224478)
(4,0.426759508)
(5,0.40040251)
(6,0.384634595)
(7,0.363593454)
(8,0.346087998)
(9,0.332107288)
(10,0.326745799)
(11,0.311575041)
(12,0.30399909)
(13,0.299667582)
(14,0.288340208)
(15,0.282442882)
(16,0.275267928)
(17,0.271501679)
(18,0.267937987)
(19,0.262266881)
(20,0.257590703)
(21,0.255599833)
(22,0.250973973)
(23,0.24776231)
(24,0.241819074)
(25,0.239341017)
(26,0.238422062)
(27,0.236478758)
(28,0.231716682)
(29,0.230772535)
(30,0.228826243)
(31,0.22882313)
(32,0.225377031)
(33,0.221251733)
(34,0.218162468)
(35,0.227031198)
(36,0.215799067)
(37,0.227760011)
(38,0.214234929)
(39,0.211140055)
(40,0.208483458)
(41,0.209055964)
(42,0.204872709)
(43,0.208044992)
(44,0.203497264)
(45,0.204845486)
(46,0.20530177)
(47,0.202094266)
(48,0.203649787)
(49,0.199258097)
(50,0.199936236)
};
\legend{Train, Validation}
\end{axis}
\end{tikzpicture}

%% file: figures/w2v_dnn.tex
\begin{tikzpicture}[scale=0.6, every node/.style={scale=1.0}]
    \begin{axis}[
        width  = 0.8*\textwidth,
        height = 8cm,
        ymin=0.0,ymax=1.12,
        ytick={0,0.2,0.4,0.6,0.8,1.0},
        major x tick style = transparent,
        ybar=5*\pgflinewidth,
        bar width=11.0pt,
        xlabel = {Embedding vector length},
        ylabel = {Accuracy},
        symbolic x coords={$N=2$,$N=31$,$N=100$},
	y tick label style={
    		/pgf/number format/.cd,
   		fixed,
   		fixed zerofill,
    		precision=2},
        xtick={$N=2$,$N=31$,$N=100$},
        x tick label style={
		font=\small,
		},
        nodes near coords,
        every node near coord/.append style={rotate=90, 
        								   anchor=west,
								   font=\footnotesize,
								   /pgf/number format/.cd,
								   	fixed zerofill,
									precision=2
								   },
        enlarge x limits=0.25,
        legend cell align=left,
        legend pos=south east,
    ]
\addplot[fill=red,opacity=1.00] 
coordinates {
($N=2$,0.90)
($N=31$,0.94)
($N=100$,0.94)
};
\addplot[fill=blue,opacity=1.00] 
coordinates {
($N=2$,0.91)
($N=31$,0.93)
($N=100$,0.95)
};
\addplot[fill=green,opacity=1.00] 
coordinates {
($N=2$,0.92)
($N=31$,0.94)
($N=100$,0.94)
};
\addplot[fill=yellow,opacity=1.00] 
coordinates {
($N=2$,0.91)
($N=31$,0.94)
($N=100$,0.94)
};
\addplot[fill=brown,opacity=1.00] 
coordinates {
($N=2$,0.90)
($N=31$,0.93)
($N=100$,0.93)
};
\legend{$W=1$,$W=5$,$W=10$,$W=30$,$W=100$}
\end{axis}
\end{tikzpicture}

%% file: figures/conf_w2v_dnn.tex
\begin{tikzpicture}[scale=0.5]
    \begin{axis}[
        width=10cm,
        height=10cm,
	colormap={bluewhite}{color=(white) rgb255=(100,149,237)},
        xticklabels={BHO,OnLineGames,Renos,CeeInject,FakeRean,Vobfus,Winwebsec},
        xtick={0,...,6},
        xtick style={draw=none},
	xticklabel style={anchor=east,rotate=60,yshift=-5pt},
        yticklabels={BHO,OnLineGames,Renos,CeeInject,FakeRean,Vobfus,Winwebsec},
        ytick={0,...,6},
        ytick style={draw=none},
        enlargelimits=false,
        colorbar,
        colorbar style={
            ytick={0.0,0.2,0.4,0.6,0.8,1.0},
            yticklabels={0.0,0.2,0.4,0.6,0.8,1.0},
            yticklabel={\pgfmathprintnumber\tick},
            yticklabel style={
            		/pgf/number format/fixed,
			/pgf/number format/precision=1}
        },
        point meta min=0.0,
        point meta max=1.0,
        nodes near coords={\pgfmathprintnumber\pgfplotspointmeta},
        nodes near coords black white/.style={
            small value/.style={
                yshift=-7pt,
                text=black,
                /pgf/number format/fixed,
                /pgf/number format/precision=3
            },
            large value/.style={
                yshift=-7pt,
                text=white,
                /pgf/number format/fixed,
                /pgf/number format/precision=3
            },
            every node near coord/.style={
                check for zero/.code={
                    \pgfmathfloatifflags{\pgfplotspointmeta}{0}{
                        \pgfkeys{/tikz/coordinate}
                    }{
                        \begingroup
                        \pgfkeys{/pgf/fpu}
                        \pgfmathparse{\pgfplotspointmeta<#1}
                        \global\let\result=\pgfmathresult
                        \endgroup
                        %
                        %
                        \pgfmathfloatcreate{1}{1.0}{0}
                        \let\ONE=\pgfmathresult
                        \ifx\result\ONE
                            \pgfkeysalso{/pgfplots/small value}
                        \else
                            \pgfkeysalso{/pgfplots/large value}
                        \fi
                    }
                },
                check for zero,
            },
        },
        nodes near coords black white=0.5,
    ]
        \addplot[
            matrix plot,
            mesh/cols=7,
            point meta=explicit,draw=gray
        ] table [meta=C] {
            x y C
 0  0 0.959
 1  0 0.010
 2  0 0.010
 3  0 0.000
 4  0 0.010
 5  0 0.000
 6  0 0.010
 0  1 0.000
 1  1 0.947
 2  1 0.000
 3  1 0.032
 4  1 0.011
 5  1 0.000
 6  1 0.011
 0  2 0.000
 1  2 0.020
 2  2 0.911
 3  2 0.000
 4  2 0.020
 5  2 0.030
 6  2 0.020
 0  3 0.000
 1  3 0.020
 2  3 0.000
 3  3 0.940
 4  3 0.010
 5  3 0.010
 6  3 0.020
 0  4 0.000
 1  4 0.030
 2  4 0.000
 3  4 0.010
 4  4 0.911
 5  4 0.000
 6  4 0.050
 0  5 0.000
 1  5 0.000
 2  5 0.000
 3  5 0.000
 4  5 0.000
 5  5 1.000
 6  5 0.000
 0  6 0.000
 1  6 0.000
 2  6 0.000
 3  6 0.020
 4  6 0.050
 5  6 0.000
 6  6 0.931
        };
    \end{axis}
\end{tikzpicture}
%

%% file: figures/robust_hmm.tex
\begin{tikzpicture}[scale=0.525]
\begin{axis}[smooth,
		   width=0.5\textwidth,
		   height=0.45\textwidth,
	 	   x tick label style={
   		 	/pgf/number format/.cd,
			/pgf/number format/1000 sep={},
   			fixed,
   			fixed zerofill,
    			precision=0
		   },
	 	   y tick label style={
    		 	/pgf/number format/.cd,
   			fixed,
   			fixed zerofill,
    			precision=1
		    },
                    xmin=0,xmax=40,
                    ymin=-0.02,ymax=1.02,
                    legend pos=south west,
                    legend cell align={left},
                    legend style={font=\small},
                    xtick={0,10,20,30,40},
                    ytick={0.0,0.2,0.4,0.6,0.8,1.0},
                    xlabel={Scrambling percentage},
                    ylabel={Accuracy},
                    ylabel shift = -2.5pt] 
\addplot[color=red,thick,
no marks] coordinates {
(0,0.88)
(10,0.39)
(20,0.37)
(30,0.33)
(40,0.325)
};
\addplot[color=green,thick,
no marks] coordinates {
(0,0.849)
(10,0.5266666667)
(20,0.5133333333)
(30,0.49)
(40,0.485)
};
\addplot[color=orange,thick,
no marks] coordinates {
(0,0.96)
(10,0.5766666667)
(20,0.595)
(30,0.5683333333)
(40,0.55)
};
\addplot[color=blue,thick,
no marks] coordinates {
(0,0.9599999785)
(10,0.6499999762)
(20,0.676666677)
(30,0.6733333468)
(40,0.6616666913)
};
\legend{HMM2Vec-SVM, HMM2Vec-$k$NN, HMM2Vec-RF, HMM2Vec-DNN}
\end{axis}
\end{tikzpicture}

%% file: figures/robust_w2v.tex
\begin{tikzpicture}[scale=0.525]
\begin{axis}[smooth,
		   width=0.5\textwidth,
		   height=0.45\textwidth,
	 	   x tick label style={
   		 	/pgf/number format/.cd,
			/pgf/number format/1000 sep={},
   			fixed,
   			fixed zerofill,
    			precision=0
		   },
	 	   y tick label style={
    		 	/pgf/number format/.cd,
   			fixed,
   			fixed zerofill,
    			precision=1
		    },
                    xmin=0,xmax=40,
                    ymin=-0.02,ymax=1.02,
                    legend pos=south west,
                    legend cell align={left},
                    legend style={font=\small},
                    xtick={0,10,20,30,40},
                    ytick={0.0,0.2,0.4,0.6,0.8,1.0},
                    xlabel={Scrambling percentage},
                    ylabel={Accuracy},
                    ylabel shift = -2.5pt] 
\addplot[color=red,thick,
no marks] coordinates {
(0,1.00)
(10,0.6866666667)
(20,0.6666666667)
(30,0.65)
(40,0.655)
};
\addplot[color=green,thick,
no marks] coordinates {
(0,1.00)
(10,0.6633333333)
(20,0.660)
(30,0.685)
(40,0.6616666667)
};
\addplot[color=orange,thick,
no marks] coordinates {
(0,0.980)
(10,0.6183333333)
(20,0.6133333333)
(30,0.6116666667)
(40,0.6383333333)
};
\addplot[color=blue,thick,
no marks] coordinates {
(0,0.960)
(10,0.6666666667)
(20,0.54)
(30,0.5233333333)
(40,0.5183333333)
};
\addplot[color=brown,thick,
no marks] coordinates {
(0,0.930)
(10,0.5766666667)
(20,0.6016666667)
(30,0.5816666667)
(40,0.5666666667)
};
\legend{$W=1$, $W=5$, $W=10$, $W=30$, $W=100$}
\end{axis}
\end{tikzpicture}

%% file: figures/robust_w2v_knn.tex
\begin{tikzpicture}[scale=0.525]
\begin{axis}[smooth,
		   width=0.5\textwidth,
		   height=0.45\textwidth,
	 	   x tick label style={
   		 	/pgf/number format/.cd,
			/pgf/number format/1000 sep={},
   			fixed,
   			fixed zerofill,
    			precision=0
		   },
	 	   y tick label style={
    		 	/pgf/number format/.cd,
   			fixed,
   			fixed zerofill,
    			precision=1
		    },
                    xmin=0,xmax=40,
                    ymin=-0.02,ymax=1.02,
                    legend pos=south west,
                    legend cell align={left},
                    legend style={font=\small},
                    xtick={0,10,20,30,40},
                    ytick={0.0,0.2,0.4,0.6,0.8,1.0},
                    xlabel={Scrambling percentage},
                    ylabel={Accuracy},
                    ylabel shift = -2.5pt] 
\addplot[color=red,thick,
no marks] coordinates {
(0,0.957)
(10,0.7033333333)
(20,0.6933333333)
(30,0.6183333333)
(40,0.60)
};
\addplot[color=green,thick,
no marks] coordinates {
(0,0.9505)
(10,0.635)
(20,0.7416666667)
(30,0.6783333333)
(40,0.665)
};
\addplot[color=orange,thick,
no marks] coordinates {
(0,0.9435)
(10,0.70)
(20,0.7183333333)
(30,0.6916666667)
(40,0.6833333333)
};
\addplot[color=blue,thick,
no marks] coordinates {
(0,0.931)
(10,0.7466666667)
(20,0.6233333333)
(30,0.5833333333)
(40,0.5583333333)
};
\addplot[color=brown,thick,
no marks] coordinates {
(0,0.914)
(10,0.6933333333)
(20,0.6316666667)
(30,0.5716666667)
(40,0.5816666667)
};
\legend{$W=1$, $W=5$, $W=10$, $W=30$, $W=100$}
\end{axis}
\end{tikzpicture}

%% file: figures/robust_w2v_rf.tex
\begin{tikzpicture}[scale=0.525]
\begin{axis}[smooth,
		   width=0.5\textwidth,
		   height=0.45\textwidth,
	 	   x tick label style={
   		 	/pgf/number format/.cd,
			/pgf/number format/1000 sep={},
   			fixed,
   			fixed zerofill,
    			precision=0
		   },
	 	   y tick label style={
    		 	/pgf/number format/.cd,
   			fixed,
   			fixed zerofill,
    			precision=1
		    },
                    xmin=0,xmax=40,
                    ymin=-0.02,ymax=1.02,
                    legend pos=south west,
                    legend cell align={left},
                    legend style={font=\small},
                    xtick={0,10,20,30,40},
                    ytick={0.0,0.2,0.4,0.6,0.8,1.0},
                    xlabel={Scrambling percentage},
                    ylabel={Accuracy},
                    ylabel shift = -2.5pt] 
\addplot[color=red,thick,
no marks] coordinates {
(0,0.99)
(10,0.615)
(20,0.7383333333)
(30,0.6566666667)
(40,0.6316666667)
};
\addplot[color=green,thick,
no marks] coordinates {
(0,0.99)
(10,0.6616666667)
(20,0.7316666667)
(30,0.6583333333)
(40,0.645)
};
\addplot[color=orange,thick,
no marks] coordinates {
(0,0.99)
(10,0.725)
(20,0.7166666667)
(30,0.705)
(40,0.6883333333)
};
\addplot[color=blue,thick,
no marks] coordinates {
(0,0.98)
(10,0.7416666667)
(20,0.53)
(30,0.515)
(40,0.46)
};
\addplot[color=brown,thick,
no marks] coordinates {
(0,0.97)
(10,0.7583333333)
(20,0.5616666667)
(30,0.5466666667)
(40,0.5316666667)
};
\legend{$W=1$, $W=5$, $W=10$, $W=30$, $W=100$}
\end{axis}
\end{tikzpicture}

%% file: figures/robust_w2v_dnn.tex
\begin{tikzpicture}[scale=0.525]
\begin{axis}[smooth,
		   width=0.5\textwidth,
		   height=0.45\textwidth,
	 	   x tick label style={
   		 	/pgf/number format/.cd,
			/pgf/number format/1000 sep={},
   			fixed,
   			fixed zerofill,
    			precision=0
		   },
	 	   y tick label style={
    		 	/pgf/number format/.cd,
   			fixed,
   			fixed zerofill,
    			precision=1
		    },
                    xmin=0,xmax=40,
                    ymin=-0.02,ymax=1.02,
                    legend pos=south west,
                    legend cell align={left},
                    legend style={font=\small},
                    xtick={0,10,20,30,40},
                    ytick={0.0,0.2,0.4,0.6,0.8,1.0},
                    xlabel={Scrambling percentage},
                    ylabel={Accuracy},
                    ylabel shift = -2.5pt] 
\addplot[color=red,thick,
no marks] coordinates {
(0,1.00)
(10,0.6399999857)
(20,0.6733333468)
(30,0.6283333302)
(40,0.6100000143)
};
\addplot[color=green,thick,
no marks] coordinates {
(0,1.00)
(10,0.6483333111)
(20,0.6733333468)
(30,0.6666666865)
(40,0.6499999762)
};
\addplot[color=orange,thick,
no marks] coordinates {
(0,1.00)
(10,0.6333333254)
(20,0.6133333445)
(30,0.6399999857)
(40,0.6316666603)
};
\addplot[color=blue,thick,
no marks] coordinates {
(0,0.9666666389)
(10,0.7883333564)
(20,0.6050000191)
(30,0.6133333445)
(40,0.5416666865)
};
\addplot[color=brown,thick,
no marks] coordinates {
(0,0.9166666865)
(10,0.7549999952)
(20,0.6100000143)
(30,0.5883333087)
(40,0.5799999833)
};
\legend{$W=1$, $W=5$, $W=10$, $W=30$, $W=100$}
\end{axis}
\end{tikzpicture}

%% file: figures/best.tex
\begin{tikzpicture}[scale=0.6, every node/.style={scale=1.0}]
    \begin{axis}[
        width  = 0.75*\textwidth,
        height = 7.5cm,
        ymin=0,ymax=1.14,
        ytick={0.0,0.2,0.4,0.6,0.8,1.0},
        major x tick style = transparent,
        ybar=5*\pgflinewidth,
        bar width=18pt,
        xlabel = {Learning technique},
        ylabel = {Accuracy},
        xlabel shift = 6pt,
        symbolic x coords={
HMM2Vec-SVM,
HMM2Vec-$k$NN,
HMM2Vec-RF,
HMM2Vec-DNN,
Word2Vec-SVM,
Word2Vec-$k$NN,
Word2Vec-RF,
Word2Vec-DNN
	},
        xtick={
HMM2Vec-SVM,
HMM2Vec-$k$NN,
HMM2Vec-RF,
HMM2Vec-DNN,
Word2Vec-SVM,
Word2Vec-$k$NN,
Word2Vec-RF,
Word2Vec-DNN
	},
	y tick label style={
		font=\small,
    		/pgf/number format/.cd,
   		fixed,
   		fixed zerofill,
		1000 sep={},
    		precision=2},
        x tick label style={
        		rotate=60,
		font=\footnotesize,
		anchor=north east,
		inner sep=0mm
		},
        nodes near coords,
        every node near coord/.append style={rotate=90, 
        								   anchor=west,
								   font=\footnotesize,
								   /pgf/number format/.cd,
								   	fixed zerofill,
									1000 sep={},
									precision=2
								   },
        enlarge x limits=0.08,
    ]
\addplot[fill=blue,opacity=1.00] 
coordinates {
(HMM2Vec-SVM,0.89)
(HMM2Vec-$k$NN,0.79)
(HMM2Vec-RF,0.96)
(HMM2Vec-DNN,0.94)
(Word2Vec-SVM,0.95)
(Word2Vec-$k$NN,0.89)
(Word2Vec-RF,0.96)
(Word2Vec-DNN,0.94)
};
\end{axis}
\end{tikzpicture}